\documentclass{emulateapj}
\usepackage{natbib}
\usepackage{times}
\usepackage{graphicx}
\usepackage{amsmath}
\usepackage{hyperref}

\newcommand{\Ltil}{\tilde{L}}
\newcommand{\rtil}{\tilde{r}}
\newcommand{\Mtil}{\tilde{M}}
\newcommand{\Mutil}{\tilde{\mathcal{M}}}
\newcommand{\rhalf}{r_{1/2}}
\newcommand{\Mhalf}{M_{1/2}}
\newcommand{\Lhalf}{L_{1/2}}

\newcommand{\movieurl}{\url{http://www.physics.uci.edu/~bullock/fcurve/movies.html}}

\begin{document}

\title{From Galaxy Clusters to Ultra-Faint Dwarf Spheroidals: A Fundamental Curve Connecting Dispersion-supported Galaxies to Their Dark Matter Halos}
\shorttitle{Fundamental Curve}

\author{Erik J. Tollerud\altaffilmark{1}, James S. Bullock\altaffilmark{1}, Genevieve J. Graves\altaffilmark{2}, Joe Wolf\altaffilmark{1}}
\altaffiltext{1}{Center for Cosmology, Department of Physics and Astronomy, 4129 Frederick Reines Hall, University of California, Irvine, CA, 92697, USA; etolleru@uci.edu, bullock@uci.edu, wolfj@uci.edu}
\altaffiltext{2}{Miller Fellow, Department of Astronomy, 601 Campbell Hall, University of California, Berkeley, CA 94720; graves@astro.berkeley.edu}

\begin{abstract}
We examine scaling relations of dispersion-supported galaxies over more than eight orders of magnitude in luminosity
by transforming standard fundamental plane parameters into a space of mass, radius, and luminosity.  
The radius variable $r_{1/2}$ is the de-projected (3-D) half-light radius, the mass variable $M_{1/2}$ is  the total gravitating mass within this radius, and $L_{1/2}$ is half the luminosity.     
We find that from ultra-faint dwarf spheroidals to giant cluster spheroids, dispersion-supported galaxies scatter about a one-dimensional ``fundamental curve'' through this MRL space. The mass-radius-luminosity relation transitions from $M_{1/2} \sim r_{1/2}^{1.44} \sim L_{1/2}^{0.30}$  for the faintest dwarf spheroidal galaxies  to $M_{1/2} \sim r_{1/2}^{1.42} \sim L_{1/2}^{3.2}$ for the most
luminous galaxy cluster spheroids. The weakness of the $M_{1/2}-L_{1/2}$ slope on the faint end may imply that potential well depth limits galaxy formation in small galaxies, while the stronger dependence on $L_{1/2}$ on the bright end suggests that baryonic physics limits galaxy formation in massive galaxies.
The mass-radius projection of this curve can be compared to median dark matter halo mass profiles of $\Lambda$CDM halos  in order to construct a virial mass-luminosity relationship ($M_{\rm vir} - L$) 
for galaxies that spans seven orders of magnitude in $M_{\rm vir}$.    
Independent of any global abundance or clustering information, we find that (spheroidal) galaxy formation needs to be most efficient in halos of $M_{\rm vir} \sim 10^{12} \, M_\odot$ and to become inefficient above and below this scale.  
Moreover, this profile matching 
technique for deriving the $M_{\rm vir} - L$ is most accurate at the high and low luminosity extremes (where dark matter fractions are highest) and is therefore quite complementary to statistical approaches that rely on having a well-sampled luminosity function.   We also consider the significance and utility of the scatter about this relation, and find that in the dSph regime observational errors are almost at the point where we can explore the intrinsic scatter in the luminosity-virial mass relation.  
Finally, we note that purely stellar systems like Globular Clusters and Ultra Compact Dwarfs do not follow the fundamental curve relation.  This allows them to easily be distinguished from dark-matter dominated dSph  galaxies in MRL space.
\end{abstract}

\section{Introduction}
\label{sec:intro}
Galaxy observables such as size, luminosity, and velocity dispersion 
are known to follow scaling relations.  The study of these relations provides 
a window  into the processes that regulate galaxy formation.
 The $\Lambda$CDM dark matter halos that host these galaxies are also predicted to follow
 structural scaling relations, including relations between their central densities and total virial masses.
 In this paper, we seek to link galaxy observables to dark matter halo properties by
  studying galaxy dynamical masses ($M_{1/2}$) within their 3-D half-light radii ($r_{1/2}$) as a 
 function of galaxy luminosity ($L_{1/2} = L/2$).  This coordinate space of intrinsic parameters (MRL Space)
 is obtained via a simple transformation of the standard {\em observed} parameters of fundamental plane space.
Our approach is motivated by the work of \citet{wolf09}, who showed that the dynamical 
 mass of a spheroidal galaxy within $r_{1/2}$ 
 can be determined accurately from observed sizes and velocity dispersions without knowledge
 of the stellar velocity dispersion anisotropy.  This fact enables  manifestly apparent physical interpretations of MRL space
 and,  in principle, a method to connect central galaxy densities to global dark matter halo properties.
   
It is well established that when placed in a parameter space of observed velocity dispersion ($\sigma$), 2-D effective radius ($R_e$), and surface brightness ($I_e$), bright ($\gtrsim L_*$) early-type galaxies lie approximately within a two-dimensional ``fundamental plane'' \citep{dj87fp,dressler87dnsig,faber87fp}. Other work \citep[e.g.][]{nieto90faintfp,b2f1,burbend97,pb02fundline,zar06fman,shankar06,woo08,forbes08} has expanded upon or considered similar such relations sometimes including galaxies that have significant rotationally-supported components.  These scaling relations provide a wealth of opportunities to examine what physical processes generate them \citep[e.g.][]{dantas00,dekel03fundline,robertson06ellscale,zar08eqgal,hopkins08fpdiss,korm09ESph,bovill09,graves09ii}, and hence further constrain scenarios of galaxy formation.

Zarisky and collaborators \citep*{zar06fmdwarfs,zar06fman} explored a unified description of the fundamental plane parameters 
for all spheroids that are embedded within their own dark matter halos.  
They found that dwarf spheroidal galaxies (dSphs), dwarf elliptical galaxies (dE), normal elliptical galaxies (E), and
the extended  stellar spheroidal components of  galaxy clusters (cluster spheroids, CSphs)
could be characterized by a 2-D fundamental manifold in ($\sigma$, $R_e$,  $I_e$) space (although curved relations had been seen noted in sub-spaces, e.g. \citealt{gra05}).   
 The CSph of a galaxy cluster halo is the sum of the brightest cluster galaxy (BCG) and
the extended intra-cluster stars (ICS).   Empirically, the inclusion of CSphs is motivated by the fact that they demonstrate a relationship between $R_e$ and $I_e$ that is similar to elliptical galaxies in many respects \citep{gonz05icl}.  From a theoretical/cosmological perspective, a CSph is the most natural single stellar system to associate with the host dark matter halo of a cluster
\citep[while the cluster galaxies themselves are more readily associated with subhalos, e.g.][]{conroy07,purcell07icl}.
Typically, the CSph ($L \gtrsim 10^{11} L_\odot$)
contains many more stars than the BCG by itself \citep{gonz05icl} and, importantly for our purposes, it 
extends to a fair fraction of the cluster virial radius, and thus (in principle) allows a more global probe of the 
cluster potential.  We will include CSphs as the cluster-halo counterparts to normal spheroidal galaxies in our work. 
Hence, our definition of ``galaxy'' here is the central luminous component of a distinct dark matter halo (although it may be a subhalo of a 
larger halo, as is the case for dSph satellites or non-BCG galaxies in clusters).

    On the opposite luminosity/size extreme from giant CSphs are the ultra-faint dwarf dSphs \citep[e.g.][]{will05wI,bel07catsdogs,walsh07boo,bel2009seg2}.  Discovered by searches within the Sloan Digital Sky Survey dataset \citep[SDSS,][]{sdssdr6}, these systems can have luminosities smaller than  $\sim 1000 L_\odot$
and have been shown to be the most dark matter dominated systems known
\citep[e.g.][]{martin07,sandg07,pen08dwarfdm,geha09seg1,simon10seg1}.  The ultra-faint dSphs provide means to study galaxy formation within the smallest dark matter halos that host stars \citep{stri08dmdom}.  By including them in our analysis, we will extend galaxy scaling relation studies to span more than eight orders of magnitude in luminosity.

By exploring galaxy properties over a very wide range in luminosity, we are able to address one of the
broader questions in astrophysics: how and why galaxy formation efficiency varies as a function
of dark matter halo mass.  
 Remarkably, observed galaxy luminosity functions and two-point clustering statistics
can be explained fairly well under the assumption that $L$ (or stellar mass) 
maps to dark matter halo mass
 $M_{\rm vir}$ in a monotonic way
\citep[e.g.][]{krav04hod,CW08,moster09abund}, such that halo viral mass-to-light ratio $M_{\rm vir}/L$ is 
minimized near $L \sim L_*$ and rises steeply at larger and smaller $L$.  An understanding of this behavior 
-- how and why it happens -- is hampered at the smallest and largest mass scales because luminosity functions
become less complete and less well-sampled in the extremes.
One of the goals of this work is to use galaxy MRL relations to inform the $M_{\rm vir} - L$ 
mapping in a way that is independent of large-scale abundance and clustering studies.

The paper is organized as follows: in \S \ref{sec:dat}, we describe the data set used for this study and the relevant sources.  In \S \ref{sec:mlrspace} we consider the scaling relations of our data, introducing a new space (``MRL'' Space) that is designed to provide a bridge between the scaling relations of galaxies and the scaling relations of dark matter halos over the full dynamic range of known galaxies ($\sim10$ orders of magnitude in $L$ and $M_{\rm vir}$).  In \S \ref{sec:curve}, we introduce a one-dimensional curve that the galaxies follow in this space, and apply this curve to canonical $\Lambda$CDM halos to map halos onto their galaxies. In \S \ref{sec:err} we address the scatter in the fundamental curve, in \S \ref{sec:err2} we address the errors and scatter in the halo mapping, and in \S \ref{sec:conc}, we conclude.

Throughout this paper we assume a $\Lambda$CDM cosmology with WMAP7 \citep{WMAP7} parameters of $h=0.704$, $\Omega_M=.272$, $\Omega_\Lambda=1-\Omega_M$, $\sigma_8=0.809$, and $n_s=0.963$.  Further, we use the symbol $\log$ to represent base-10 logarithms.

\begin{deluxetable*}{cccccccccc}[t!]
\tablecolumns{10}
\tablecaption{Key properties of dispersion-supported stellar systems.}
\tablehead{
\colhead{Name \tablenotemark{a}} &
\colhead{$\log(\sigma)$ \tablenotemark{b}} &
\colhead{$\log(R_e)$ \tablenotemark{c}} &
\colhead{$\log(r_{1/2})$ \tablenotemark{d}} &
\colhead{$\log(L_{1/2})$ \tablenotemark{e}} &
\colhead{$\log(M_{1/2})$ \tablenotemark{f}} &
\colhead{$\log(M_{1/2}^{\rm DM})$ \tablenotemark{g}} &
\colhead{Object Type \tablenotemark{h}} &
\colhead{$N_{\rm obj}$ \tablenotemark{i}} &
\colhead{Source \tablenotemark{j}} 
}

\startdata
A0122 & 2.83 & 2.03 & 2.15 & 11.3 & 13.7 & 13.7 &CSph & \nodata & 1 \\
E Bin 1 & 1.92 & 0.04 & 0.16 & 9.22 & 9.85 & 9.5 & E & 36 & 2 \\
VCC452 & 1.38 & -0.15 & -0.02 & 8.04 & 8.57 & 8.37 & dE & \nodata & 3 \\
Draco & 1.0 & -0.66 & -0.53 & 5.03 & 7.32 & 7.32 & dSph & \nodata & 4 \\
47 Tuc & 1.31 & -2.66 & -2.54 & 5.20 & 5.92 & 5.92 & GC & \nodata & 5,6 \\ 
F-19 & 1.36 & -1.05 & -0.92 & 7.00 & 7.64 & 7.64 & UCD & \nodata & 7

\enddata

\tablenotetext{a}{Name of the object.}
\tablenotetext{b}{Log of Velocity dispersion in km/s.}
\tablenotetext{c}{Log of 2-D Half-light/effective radius in kpc.}
\tablenotetext{d}{Log of 3-D (deprojected) Half-light radius in kpc. (see \S \ref{sec:mlrspace})}
\tablenotetext{e}{Log of V-band half-luminosity in $L_\odot$, i.e. $L_V/2$.}
\tablenotetext{f}{Log of Half-light mass from Equation \ref{eqn:Mh} in $M_\odot$.}
\tablenotetext{g}{Log of Corrected/dark half-light mass in $M_\odot$ as described in \S \ref{sec:curve}.}
\tablenotetext{h}{Object Type: CSph = Cluster Spheroid as described in \citet{zar06fman}, E=bright Elliptical galaxy, dE=dwarf Elliptical, dSph=local group dwarf spheroidal, GC=Galactic globular cluster, UCD=Ultra Compact Dwarf.}
\tablenotetext{i}{Number of objects per bin for E data set.}
\tablenotetext{j}{Source code: 1)\citet{zar06fman}, 2)\citet{graves09ii}, 3)\citet{geha03deii},  4)\citet{wolf09}, 5)\citet{harris96gcs}, 6)\citet{PM93}, 7)\citet{mieske08ucds}.}

\tablecomments{This table will be published in its entirety in the electronic edition of the \emph{Journal}. A portion is shown here for guidance regarding form and content.}

\label{tab:dat}
\end{deluxetable*}

\section{Data}
\label{sec:dat}
The data sources for this study are varied by necessity due to the wide dynamic range covered.  Table \ref{tab:dat} gives the relevant parameters for the objects in this study and the sources for each.  Starting with the least luminous objects that are embedded within dark matter halos ($L \lesssim 10^{8} L_\odot$),
our dwarf spheroidal (dSph) data set is taken from the summary table of \citet{wolf09} and draws from
various sources for photometric properties and resolved star kinematic measurements for Milky Way dSph galaxies.
Moving up in brightness ($L \simeq 10^{8-9} L_\odot$)   our ``dwarf elliptical'' (dE) sample is
taken from the Virgo Cluster dE study of \citet{geha03deii}. 
Note that while dEs are not as clearly dark matter dominated as dSph galaxies within $r_{1/2}$ (see below), they
are believed to be embedded in their own dark matter halos based on extended kinematic samples \citep{geha10}. 

Data for normal elliptical galaxies  (E)  are from \citet[][$L \simeq 10^{10} L_\odot$]{graves09i} and are discussed in more detail toward the end of this section.   The brightest ($L \simeq 10^{11} L_\odot$) cluster spheroid (CSph) data are from the imaging of \citet{gonz05icl} and spectra of \citet{zar06fman}.  These data are also described in more detail below.  

We also examine two comparison populations as examples of systems
that are not embedded within dark matter halos: Milky Way globular clusters (GCs, $L\simeq 10^{5}$) and ultra-compact dwarfs (UCDs, $L \simeq 10^{6}$).  
For GC photometry we use the 2003 revision of the Harris catalog \citep{harris96gcs} and take velocity dispersions from \citet{PM93}. For UCDs we use data from \citet{mieske08ucds}.  Note that while the status of UCDs as large examples of purely stellar systems is debated \citep[e.g.][ and references therein]{evst07ucd,goerdt08ucds,baum08ucds,dab09ucds,taylor10gcucds} we find that their scaling relations are more in
line with GCs than similarly luminous dSphs and therefore treat them as lacking dark matter halos below.

The CSph data set stands out compared to the other data sets in two distinct ways.  First, 
while all other data sources are in the V band, the CSph data \citep[summarized by][]{zar06fman} use Cousins I-band luminosities.  We convert these data to V-band using averaged colors of E galaxies from \citet{fuk95}.  While this does not account for the possibility of a systematic error in $R_e$ for the CSph data points due to a different choice of band, this effect is likely to be small given the large dynamic range in this data set.  Furthermore, \citet{labarbera08} find that the fundamental plane for early-type galaxies is nearly independent of band from optical to the K-band.  Given the similarity of the stellar populations for those galaxies and the CSph, it is therefore likely that the band mismatch is not a significant effect.     
 
The second way that the CSph data set differs is that the velocity dispersions from \citep{zar06fman} 
are derived from galaxies in the cluster, rather than the CSph (mostly ICS) light itself.  This is of course not ideal, 
but the measurement of ICS velocity dispersions is very difficult with current spectroscopic capabilities.  While this has been accomplished both in integrated light \citep{kel02} and planetary nebula kinematics \citep[e.g.][]{arn04} for a few clusters, there is not yet a large, homogenous sample.  This is required for generality and to compare to our other samples, and hence we are forced to use galaxy dispersions until large direct measurement samples become available.
In principle this could impose a bias in our mass estimator (described below) because the ICS and cluster galaxies follow different distribution functions.  We explore in more detail how this bias might affect our results in \S \ref{sec:err}.

The normal elliptical galaxy data comprise a sample of $\sim$16,000
galaxies selected from the Sloan Digital Sky Survey (SDSS,
\citealt{york00sdss}) Main Galaxy Sample \citep{strauss02sdss}, as described
in \citet{graves09i}.  Galaxies are selected to be passively
evolving quiescent objects with no emission lines in their spectra.
The individual galaxies are sorted into bins in the 3-D Fundamental
Plane parameter space defined by $\sigma$, $R_e$, and $I_e$.  Values
reported here are the median values for each bin of galaxies.

Before continuing, we summarize our galaxy terminology  and the symbol codes we use when presenting each galaxy
type. The CSph population of \citet{zar06fman} is presented as orange squares.  The ``E'' or ``bright E'' terminology refers to the \citet{graves09i} data set and is represented as red circles of varying size such that the size of the data point is proportional to the number of galaxies in each bin.  The dE or ``dwarf elliptical" label refers to the   \citet{geha03deii} data set and is presented as yellow diamonds of uniform size.  The Milky Way dSph satellites here are represented by magenta triangles.  In some cases, a distinction will be drawn between the ``SDSS dSphs'' and the ``classical dSphs,'' referring to those discovered by SDSS and those known before.  The SDSS dwarfs are almost exclusively fainter, and include the ``ultra-faint dSphs.''    Finally, the GC and UCD populations are represented by the green and blue star-symbols and pentagons, respectively.

\section{MRL Space}
\label{sec:mlrspace}

We now examine the data set described in the previous section in the context of the scaling relations of the observables. We emphasize the use of the MRL space described below to understand this data set.

\begin{figure}[htb!]
\epsscale{1.15}
\plotone{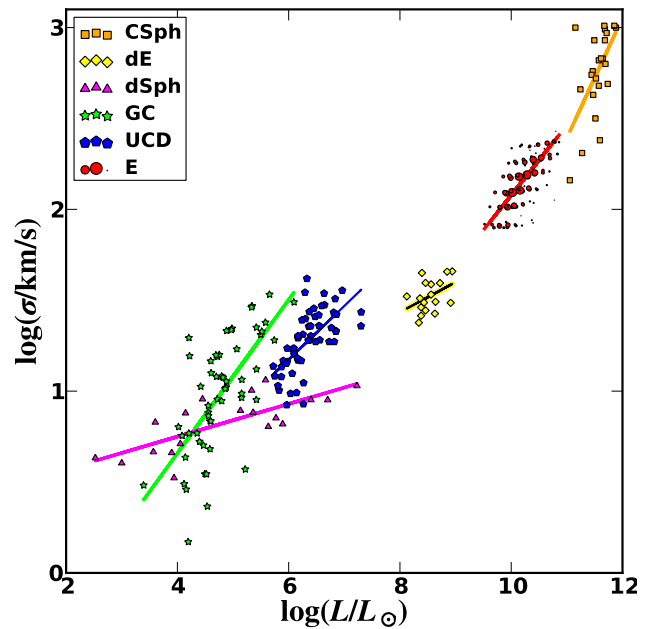}
\caption{Faber-Jackson \citep{fjrel} relation relating luminosity to stellar velocity dispersion for the data set.  The lines show best-fit power laws for each of the sub-populations of objects (see text for slopes).   Orange squares are cluster spheroids (CSph),  red circles are elliptical galaxies (E), yellow diamonds are dwarf ellipticals (dE), magenta triangles are Milky Way dwarf spheroidals (dSph), green stars are Galactic globular clusters (GC), and blue pentagons are ultra-compact dwarfs (UCD).}
\label{fig:fjplot}
\end{figure}

First, we provide a sample projection of the data set described in the previous section (Table \ref{tab:dat}).  Figure \ref{fig:fjplot} plots this data set in the 2-D space of luminosity ($L$) and stellar velocity dispersion dispersion ($\sigma$)---the Faber-Jackson relation \citep{fjrel}.  We also show best-fit power laws ($L=\sigma^{\gamma}$) for each of our classes of objects.  We compute slopes by fitting a linear relation in log space with $\log L$ ($\log \sigma$) as the parametric variable.  For the CSphs, Es, dEs, dSphs, UCDs, and GCs, this results in slopes of $\gamma=$ 1.5 (0.5), 2.6 (1.8), 6.0 (1.1), 11.1 (6.3), 2.4 (1.2), and 3.4 (1.5), respectively.  

In this plane the slopes increase towards larger luminosities, suggesting a definite scaling relation (the original Faber-Jackson relation).  We note, however, that the dSphs, UCDs, and GCs are mixed together in this projection, a clear drawback from interpreting these objects in this space.  Further, there is structure to the E sample not fully aligned with this 2-D parameter space.  The structure here is the fundamental plane \citep{dj87fp,dressler87dnsig,faber87fp} for E galaxies, distinguished from the Faber-Jackson relation by being a 3-D parameter space with the inclusion of the effective radius ($R_e$, the radius enclosing half the total luminosity) and use of mean surface brightness $I_e = L/(2 \pi R_e^2)$ in place of the luminosity.  In Appendix \ref{apx:altdata} we show this data set in the fundamental plane space (and the related $\kappa$ space of \citealt{b2f1}) for reference and comparison, but here we emphasize the use of a different parameter space, described below.

While the fundamental plane is a valuable parameter space of observables, the connection to this space from typical dark matter scaling relations is non-trivial.   In order to facilitate manifestly apparent theoretical interpretations, we introduce a set of physical variables -- a mass, a size, and a luminosity -- that are derived from the same observables.  Hence, we call this space ``MRL Space'' for the three variables:
\begin{enumerate}
	\item The half-light mass $M_{1/2} \equiv  M(<r_{1/2})$ -- the total dynamical mass within $r_{1/2}$.
	\item The 3-D half-light radius $r_{1/2}$, the radius enclosing the half-luminosity $L_{1/2}$ .
	\item The half-luminosity, $L_{1/2}$, half of the total luminosity emitted from the galaxy (\emph{not} necessarily the same as half the observed luminosity).
\end{enumerate}
We note that the luminosity variable here is defined in terms of the total luminous material in the galaxy, ignoring any attenuation that may occur as light propagates out of the galaxy.  Below we describe the transformation of observables used to closely approximate this space for the data set here.

A major motivation for the choice of these coordinates is the explicit use of  the mass within the 3-D half-light radius as the mass variable,
$M_{1/2} \equiv M(r_{1/2})$.  The adoption of this mass in particular is motivated by \citet{wolf09}, who showed that
while dynamical masses with $r \ll r_{1/2}$ and $r \gg r_{1/2}$ are largely unconstrained from 1-D velocity dispersion
data (due to weak constraints on the stellar velocity dispersion anisotropy), $M_{1/2}$ can be determined simply and accurately for spherical systems
without knowledge of the anisotropy:
\begin{equation}
M_{1/2} = 3 \, G^{-1} \, \sigma^2 \,  r_{1/2} \,.
\label{eqn:Mh}
\end{equation}

\citet{wolf09} showed that as long as the stellar velocity dispersion profile is fairly flat with radius, this mass estimator for $M_{1/2}$ is accurate for a wide range of light profiles, including the types of profiles used to fit all of the types of objects shown in Table \ref{tab:dat}.
Hence, for stellar systems with negligible rotational support, this formula provides a good estimate for the total dynamical mass within $r_{1/2}$ (assuming spherical symmetry).  

Note that Equation \ref{eqn:Mh} was \emph{not} derived using the virial theorem, but rather follows from the Jeans Equation.   The virial theorem provides only an integral constraint on the total mass traced by a stellar system and therefore cannot be used to infer precise masses (see \citealt{merr87} Appendix A and \citealt{wolf09} \S 2.1).   Similar estimators \citep[e.g.][]{spitzer69,ill76,cappellari06} have the same form (by dimensional analysis), but for most of these the coefficient is calibrated by examining high-quality data and assuming that mass follows light.   These calibrations are less useful for a wide variety of spheroidal galaxies  because there is no reason to expect that all spheroidal galaxy are homologous.   Equation \ref{eqn:Mh} is derived analytically rather than empirically, and shows that there is a \emph{particular} radius at which the mass is unbiased at any scale ($\approx r_{1/2}$).  Estimators that do not use this radius must have different virial coefficients as a function of scale. Further, Equation \ref{eqn:Mh} assumes neither mass-follow-light nor isotropy,  and hence is suited to the range of objects with various dark matter fractions that we consider here.

Further, we note that the approximation $r_{1/2} = 4 R_e/3$ is accurate for the light profiles of relevance in this paper.  As shown in \citet{ciotti91} and \citet{limaneto99}, deprojected spherical Sersic \citep{sersic63} profiles for a range of Sersic indicies are within a few percent of this relation, and the same is demonstrated for Plummer \citep{plummer1911} and King \citep{kingprof} profiles in \citet{spitz87plummer} and \citet{wolf09}.  The objects presented here are well fit by at least one of these profiles, motivating the use of the approximation.   We note here that these deprojections must assume spherical symmetry, like the $M_{1/2}$ estimator described above.

With these estimators chosen, the MRL space as derived from the observables consists of:
\begin{enumerate}
	\item $M_{1/2} = 3 \, G^{-1} \, \sigma^2 \,  r_{1/2} \,$.
	\item $r_{1/2} = 4 R_e/3$.
	\item $L_{1/2} = L/2 = I_e \pi R_e^2 $.
\end{enumerate}
Here $R_e$ is the 2-D (projected) half-light radius, $G$ is the gravitational constant, $\sigma$ is the stellar velocity dispersion of the galaxy, and $I_e$ is the mean surface brightness within $R_e$. We note that the observables here are the same as those used for the fundamental plane and thus this space can be viewed as a transformation of the fundamental plane space.

The use of $L/2$ as $L_{1/2}$ would be invalid in the presence of significant attenuation due to dust, but the objects described here are have very low gas fractions and hence likely have negligible attenuation.  
Thus the interpretation of $L_{1/2} = L(< r_{1/2})$ as the light emitted within $r_{1/2}$  is a reasonable one for these objects, and the above set of observable transforms relations are close approximations to the actual MRL variables.

Later, we will also consider a modified version of MRL space that we call dMRL space.  
In dMRL space,  the mass variable is $\Mhalf^{\rm DM} \equiv \Mhalf - M_{\rm baryon}(<r_{1/2})$,
the {\em dark matter} mass within $r_{1/2}$.  For our purposes, the difference between $\Mhalf^{\rm DM}$ and $M_{1/2}$ will only be substantial for E and dE galaxies, and is  obtained by subtracting out the stellar mass within the half-light radius for these galaxies (which contain negligible gas fractions); explicitly, $\Mhalf^{\rm DM} \simeq \Mhalf - M_*/2$. 
  It is important to recognize that the presence of radial gradients  in $M_*/L$ due to metallicity variation could render the 
use of our formula for $\Mhalf^{\rm DM} $ invalid by shifting the radius enclosing half the stellar mass from $r_{1/2}$.  However, as shown in \citet{smith09}, 
typical metallicity gradients for the E galaxies (for which $M_*$ is most important) are $\delta \log(Z)/\delta \log(r) \approx -0.1$.  Using this gradient
with a typical ancient (13.7 Gyr) solar metallicity stellar population from \citet{bc03}, we find $M_*/L$ shifts by 0.07 dex from $R_e$ to $0.1 R_e$.  Hence, this
is a small effect for our galaxies and we disregard it\footnote{In principle, a radial $M_*/L$ shift could be resolved by replacing $L_{1/2}$ by $M_{*1/2}$ and 
defining the appropriate $r_{*1/2}$.  However, the data quality is not sufficient to derive $M_*$ for our full sample, so we use $L$ here.}.  We return to dMRL space in the next section.

\begin{figure*}[htb!]
\epsscale{1.2}
\plotone{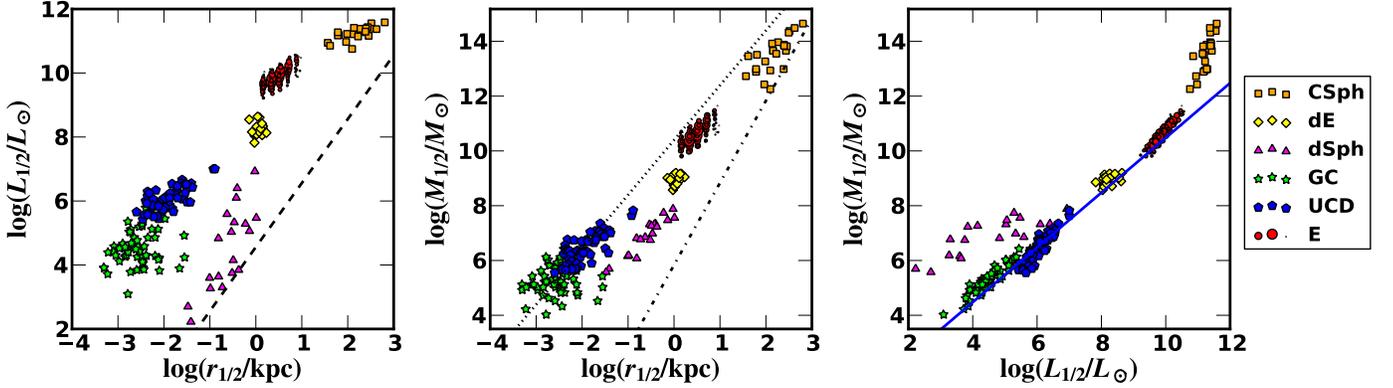}
\caption{Two dimensional projections of the three-dimensional data shown in Figure \ref{fig:fjplot}  onto the coordinate axes
of MRL space: luminosity vs. radius, mass vs. radius, and mass vs. luminosity (from left to right, respectively).
 The color/symbol code is indicated.   In the left panel, the dashed line is a constant surface brightness of $\mu_V=30$ mag/arcsec$^{2}$ -- any galaxy below this line would be undetectable with current methods.  In the middle panel, the dash-dotted line shows the scaling of constant density ($M \propto r^3$) while the dotted line shows
 the scaling of constant \emph{surface} density ($M \propto r^2$), with normalization chosen to bracket the galactic objects.   In the right panel, the solid blue line reflects
 a constant mass-to-light ratio, specifically $M_{1/2}/M_\odot=3 L_{1/2}/L_\odot$.  Slope deviations from the $M_{1/2} \propto L_{1/2}$ line are equivalent to ``tilts'' in each galaxy population's fundamental plane.  
 }
\label{fig:mlr2d}
\end{figure*}

\begin{figure*}[htb!]
\plotone{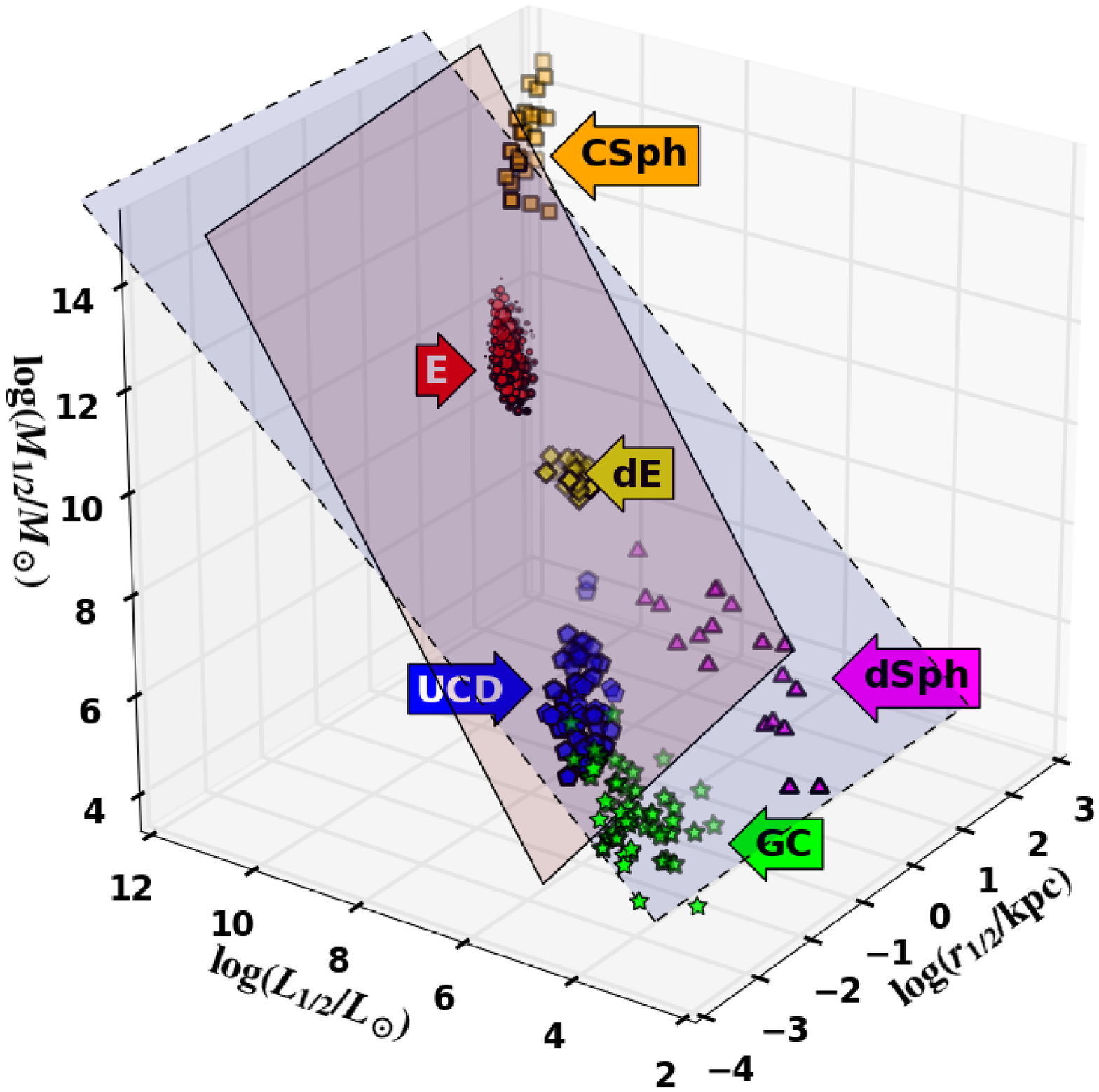}
\caption{Three dimensional representation of the MRL space of $\log(M_{1/2}/M_\odot)$, $\log(L_{1/2}/L_\odot)$, and $\log(R_{1/2}/{\rm kpc})$.  The transparent (red) plane with solid borders is the fundamental plane of \citet{graves09ii}.  The (blue) plane with dashed borders corresponds to $M_{1/2} = 3 L_{1/2}$, i.e. the plane for mass-follows-light, sometimes referred to as the ``virial plane''.   The data point color and point-type scheme matches that of the Figure \ref{fig:fjplot} (or see \S \ref{sec:dat}). 
A rotating animation of this plot is available at \movieurl.}
\label{fig:mlr3d}
\end{figure*}

In Figure \ref{fig:mlr2d} we plot the data set described in \S \ref{sec:dat} transformed into MRL space and 
projected along the coordinate axes.  For each of the projections, we have also plotted lines to reflect scalings of interest.

The left panel represents $L_{1/2}$ as a function of $r_{1/2}$.  
The galaxies show a  trend of increasing luminosity with increasing $r_{1/2}$, but occupy a relatively small fraction of the available detection space.  The GCs and UCDs, meanwhile, are much more scattered in this plot, and are consistently smaller than the dSphs at similar luminosities (e.g. higher surface brightness); as described below we interpret this (along with similar behavior in the other projections) as a clear sign they are separate populations.  The dashed black line in the left panel is a line of constant surface brightness  ($L_{1/2} \propto r_{1/2}^2$), $\mu_V=30$ mag arcsec$^{-2}$. Below this surface brightness limit, detection bias in this plane becomes significant for MW dSphs \citep{kop08,walsh09}.  This likely biases the observed $r_{1/2} - L_{1/2}$  relation to small $r_{1/2}$ at the faint end \citep{bull09stealth}.  We discuss the effect of this bias on our parameterization of the $r_{1/2} - L_{1/2}$  relation in \S \ref{subsec:mrlcurve}.

In the middle panel we show a projection into the $r_{1/2} - M_{1/2}$ space.  We include lines of constant mass density ($M_{1/2} \propto r_{1/2}^3$, black dash-dotted line) and constant \emph{surface} mass density ($M_{1/2} \propto r_{1/2}^2$, black dotted line), with normalizations arbitrarily set to bracket this data set.  In almost all cases, spherical geometry is assumed,
in which case a slope of 2,  is more properly characterized as a 3D density profile that varies as  $\rho \propto r^{-1}$ (somewhat cuspier than constant density). A slope of 3, meanwhile, is the scaling expected if all  galaxies had a single constant density within their half-light radii. This slope has been noted previously at some scales\citep{gentile09,napo10cendm,walker10dsphsp}. 
The fact that the dSph galaxies lie above the constant density line (black dash-dotted) that
is normalized to intersect the most massive cluster population suggests that they are slightly denser than galaxy clusters (but not that much denser) at their half-light
radii.  For a figure that explicitly compares the implied mean density of these objects, see Appendix \ref{apx:altdata}.
 
 Finally, in the right panel we show $M_{1/2}$ vs. $L_{1/2}$, and  a mass-follows-light line 
 ($M_{1/2} \propto L_{1/2}$) normalized at $M/L = 3$ in solar units to reflect the mass-to-light ratio of a uniform fairly old stellar population.  
 Note that the deviation of a population from  $M_{1/2} \propto L_{1/2}$ is equivalent to the ``tilt'' that is often discussed in the context of the fundamental plane. It is clear from this figure that the CSphs and dSphs deviate from this scaling substantially owing to their high dark matter fractions, while the other populations are more consistent, although the Es do show the well-known tilt, and the UCDs show a possible tilt (discussed below).
 
Figure \ref{fig:mlr3d} shows the same information, now presented in a 3-D representation.
The red plane outlined with a solid line is the  \citet{graves09ii} fundamental plane (transformed
into MRL space).  The blue plane outlined with a dashed line is a plane with mass proportional
to luminosity with $M_{1/2} = 3 L_{1/2}$ and is  indicative of the plane we would expect uniformly old, purely stellar systems
to lie within.    We note that, in fundamental plane space, this last scaling is sometimes called the ``virial plane'' (even though systems can be in virial equilibrium regardless of whether or not they lie within this plane).    In MRL space it is manifestly apparent that this plane is defined by the assumption that mass-follows-light with a fixed $M/L$.

Another feature revealed by examination of the populations in Figures \ref{fig:mlr2d} and \ref{fig:mlr3d} is a distinct separation between dSphs (magenta)
along one sequence and UCDs/GCs (blue/green) along another \citep[a similar situation is noted by][in K-band]{forbes08}.  Specifically, the UCDs and GCs cluster more closely around the 
$M_{1/2} \propto L_{1/2}$ plane (shown as dashed, transparent blue) while the dSphs (at similar luminosity) peel sharply up from it, reflecting
a significant dark matter component and larger sizes.  This difference is clearly visible in the two-dimensional projections
of MRL space shown in Figure \ref{fig:mlr2d}, and manifests itself as a wishbone-shaped bifurcation of the spheroidal sequence in Figure \ref{fig:mlr3d}.  
We also note here that the UCD sample seems to show a slight tilt from the $M_{1/2} \propto L_{1/2}$  relation, most clearly apparent in the right panel of Figure \ref{fig:mlr2d}.  This could be a sign of a very small amount of dark matter, but could also be systematic variation in the $M_*/L$ ratio due to stellar effects.  These objects have uniquely large luminosity densities, and hence are the most likely places to show changes in star formation conditions \citep{dab09ucds} or simply be an extension of scalings that exist everywhere (such as variation for the Es is described in more detail in \S \ref{sec:err}).  Alternatively, they may be due to dynamical evolution or more complex formation scenarios \citep[e.g][]{goerdt08ucds,taylor10gcucds}.  Regardless, the significance of this tilt is not clear from this data set (although more tilted than the GCs), and the UCDs and GCs are quite distinct from the dSph sample.

\begin{figure}[htbp!]
\epsscale{1.25}
\plotone{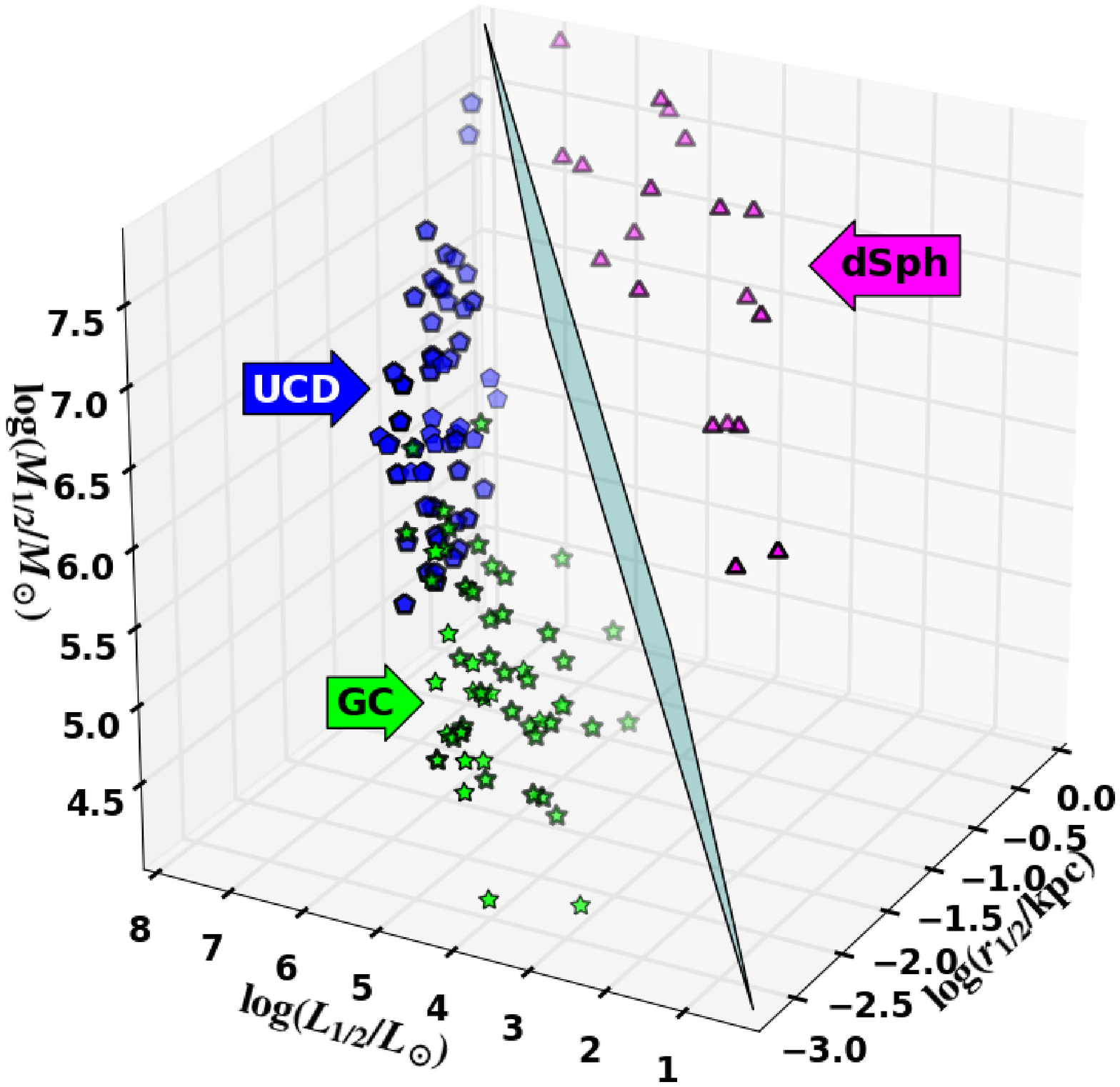}
\caption{Three dimensional representation of the dSph/GC separation plane in MRL space.  The green points are GCs, blue points are UCDs, and the yellow points are dSphs.  The transparent cyan plane is given by Equation \ref{eqn:mlrsepeq}, chosen to separate the two populations, fit using only the dSphs and the GCs (see text for details of the fit).  
A rotating animation of this plot is available at \movieurl.}
\label{fig:sepplot}
\end{figure}

Given the observation that the MW dSphs are dark matter-dominated \citep{sandg07,stri08dmdom,simon10seg1}, and GCs have $M/L$ consistent with purely stellar systems \citep[e.g.][]{PM93}, we consider if there is a clean separation between these systems based on the MRL space parameters.  We fit a plane that separates the dSphs from GCs by finding a plane that lies perpendicular to the best least squares fit between all of the dSphs and GCs and perpendicular to the best fit line through the dSph sequence; we then offset the plane until it evenly divides the two populations, giving the plane rendered in Figure \ref{fig:sepplot}.  This plane is a convenient empirical way to determine if an object is a faint dSph or a globular cluster.    In the MRL space for our data set, the best fit separation plane is given by 
\begin{equation}
\label{eqn:mlrsepeq}
0.34 \log{M_{1/2}} - 0.50 \log{L_{1/2}} + 0.79 \log{r_{1/2}} = -1.35.
\end{equation} 
Specifically, objects that lie at lower $M_{1/2}$, lower $r_{1/2}$, or higher $L_{1/2}$ are GCs while others are galaxies.  
This same relation can easily be transformed into fundamental plane space, providing the separation plane
\begin{equation}
0.68 \log{\sigma} - 0.50 \log{I_e} + 0.13 \log{R_e} = -3.23. 
\end{equation}
such that objects with lower $\sigma$, higher $\log{I_e}$, or lower $\log{R_e}$ are GCs while others are galaxies.  

The fact that this single plane easily separates the GCs and dSphs in the MRL space implies that these are distinct classes of objects (see also the discussion in Appendix \ref{apx:UCDs} - the arguments there for UCDs also apply to GCs).  It is possible that future studies of faint/low surface brightness GCs may change the location of this separation plane, or even fill in the gap, rendering the plane completely arbitrary.  But for this data set, the classes are completely separated by the plane of Figure \ref{fig:sepplot}.  Further, we note that this plane implies that a galaxy/cluster projection using a single variable \citep[e.g.][]{gil07} is not sufficient to separate these populations, as is apparent from Figure \ref{fig:mlr2d}.  All 3 dimensions are necessary to account for the most extreme objects.

Additionally, we include UCDs in Figure \ref{fig:sepplot} and find that they also lie clearly separated by the plane, even though they are \emph{not} included in the determination of the best-fit separation plane. This is suggestive that they are in the same class as GCs, and not on the galaxy sequence.  However, although given the tilt discussed above, we cannot discount the possibility that this is simply due to a relative rarity of the most massive UCDs to bridge the gap.

Given that GCs and UCDs both lack clear evidence for dark matter and sit in a distinct region of MRL space we are inclined to treat them as stellar systems rather than ``galaxies'', which we define operationally as stellar systems that are bound to a dominant dark matter halo (as discussed in \S \ref{sec:intro}).  Alternatively, a second scenario is possible where UCDs do contain significant dark matter.  If this is the case, then an interesting implication follows: there would need to be a dichotomy in galaxy formation efficiency in dark matter halos of a fixed virial mass.   Specifically, as shown in Appendix \ref{apx:UCDs}, most UCDs are consistent with no dark matter given the uncertainties in the expected stellar mass-to-light ratios.  If we force a stellar mass-to-light ratio of 2 (such that their dark matter densities are comparable to their dynamical mass densities) then the implied dark matter densities are incredibly high -- comparable to the central densities of the most massive galaxy clusters ($M_{\rm vir} \sim 10^{16} M_\odot$).
dSphs of similar luminosities sit in $M_{\rm vir} \sim 10^9 M_\odot$ halos.  UCD dark matter mass fractions would need to be extremely fined-tuned (and different from object to object) in order to avoid a dichotomy in galaxy formation efficiency at a fixed dark matter halo mass -- a dichotomy that is not seen for any other type of spheroidal system.  This is an interesting possibility and may call for more investigation, as such a result would be difficult to explain in LCDM.   

Nevertheless, we regard the above scenario to be unlikely, and adopt the simpler interpretation that UCDs are purely stellar systems that occasionally have unusually high $M_*/L$ due to unique star formation conditions or dynamical evolution.   From here on we omit the GCs and UCDs from consideration as systems that clearly contain dark matter halos of their own.   In the alternative scenario where UCDs are to be regarded as galaxies, our approach could be viewed as restricting ourselves to the simpler dSph ``branch'' of the MRL relation.

Once we remove UCDs and GCs,  we are left with a galaxy sequence in Figures \ref{fig:mlr2d} and \ref{fig:mlr3d} that scatters about a 1-D relation through MRL space.  In the next section we work towards characterizing this 1-D curve.

\begin{figure*}[htbp!]
\epsscale{1.15}
\plotone{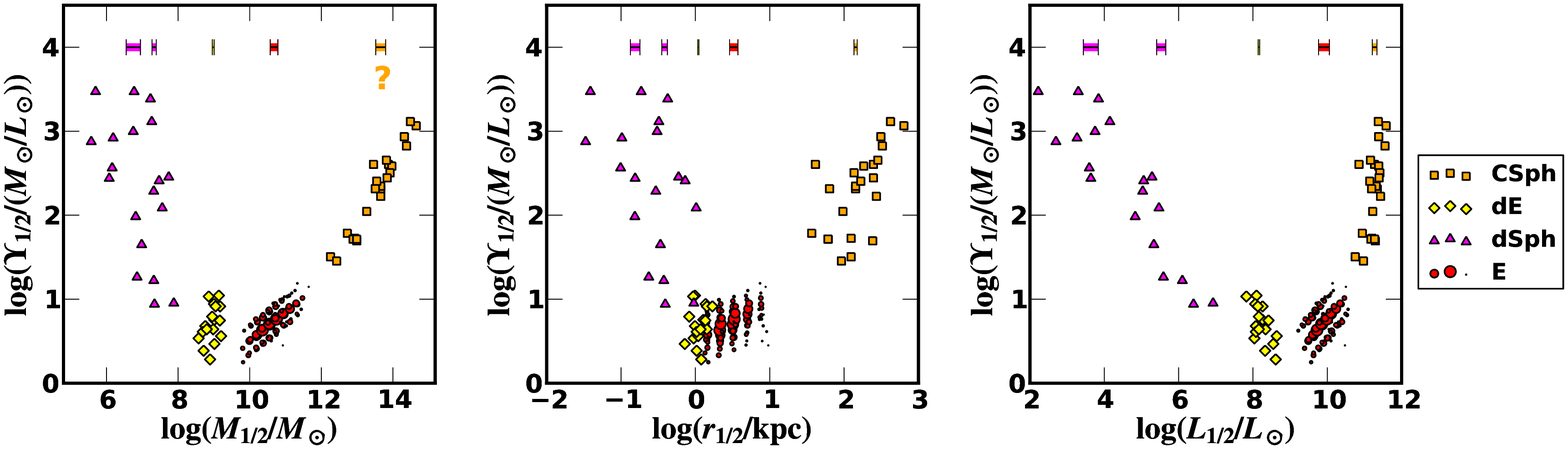}
\caption{ Mass-to-light ratios within the half-light radius, $\Upsilon_{1/2} = M_{1/2}/L_{1/2}$ in 
units of $M_\odot/L_\odot$, shown as a function of each of the MRL variables.  Error bars shown along the top of each
panel are representative of the observational uncertainties in each parameter and each galaxy type, indicated
by matching color-code and location in the M, R, or L, axis.  For the dSphs, we include separate error bars for the classical and SDSS
dwarfs. The ``?'' for the CSph $M_{1/2}$ indicates the additional (unquantifiable) uncertainty unique to the CSph, due to the use
of the cluster galaxies to determine the velocity dispersion instead of the actual dispersion of the ICS. }
\label{fig:mlr_mtol}
\end{figure*}

Figure \ref{fig:mlr_mtol} provides yet another representation of the MRL data, now presented as the dynamical 
half-light mass-to-light ratio $\Upsilon_{1/2} \equiv M_{1/2}/L_{1/2} $ (in 
$M_\odot/L_\odot$) plotted as a function of each of the MRL variables individually. 
 Along the top of each panel we show characteristic observational uncertainties for our galaxies of each type across the MRL sequence.
We discuss these errors in the context of measuring scatter in the MRL relation in \S 6.

Each panel in Figure \ref{fig:mlr_mtol} clearly reveals 
a minimum $\Upsilon_{1/2} \simeq 3$ that spans a broad regime of spheroidal galaxies, from
$M_{1/2} \simeq 10^{9-11} M_\odot$ (left); $r_{1/2} \simeq 1-10$ kpc (middle); and $L_{1/2} \simeq 10^{6-10} L_\odot$ (right).  
As discussed by \citet{wolf09} in the context of a similar figure in their paper (Figure 4), 
the  dramatic increase in dynamical half-light mass-to-light ratios at both smaller and larger scales is likely
indicative of a decrease in the efficiency of galaxy formation in the smallest and largest dark matter halos -- as discussed above, the
influence of radial variations in $M_*/L$ is $\sim 0.1$ dex, far less than that observed here. 
For the biggest, brightest, most massive galaxies, the increase in $\Upsilon_{1/2}$ implies a sharp threshold
for galaxy formation in luminosity (not in mass) at $L_{1/2} \simeq 10^{11} L_\odot$, as shown by the strong break in the left panel of Figure \ref{fig:mlr_mtol}.  
The strong sensitivity to luminosity suggests that baryonic processes are responsible for this transition.
Meanwhile, the smallest, faintest, least massive galaxies
seem to exhibit a sharp rise at a particular mass scale (not luminosity scale) near $M_{1/2} \simeq 10^6 M_\odot$ (right panel of Figure \ref{fig:mlr_mtol}).  This indicates 
they are more tied to the size of their potential wells than star formation (although this does not preclude an interaction between the two, e.g. \citealt{dekel03fundline,woo08}).
We connect these scaling trends to dark matter halo virial masses and relate them broadly to galaxy formation in Sections \ref{sec:halomatch} and \ref{sec:curve}.

\section{Fundamental Curve}
\label{sec:curve}

It is evident in Figures \ref{fig:mlr2d} and \ref{fig:mlr3d} that CSphs, Es, dEs, and dSphs seem to curve through
MRL space along a 1-D sequence \citep[see also][for dE and Es]{gra06,gra08}.  We refer to this sequence as the ``fundamental curve" and we plot analytic
representations of this curve in the left panel of Figure \ref{fig:fcurve3d} along with the associated data points.
We discuss these analytic curve representations Sections \ref{subsec:mrlcurve} and \ref{subsec:dmrlcurve}.  

It is important to note that the existence of this 1-D curve does not imply that these objects are a single parameter family, nor that the curve is a more suitable fit than a higher-dimensional construct.  As the fundamental plane \citep{graves09ii} for Es and fundamental manifold \citep{zar06fman} show, galaxies do show systematic variation along multiple directions in fundamental plane or MRL space.  We do not aim to compare the statistical significance of these relations to the fundamental curve, as the applications of 1-D and 2-D relations are quite different.  Instead, the best way to think of the fundamental curve is as the direction of largest variation of this set of dispersion-supported galaxy properties.  Thus, it is useful as the first-order scaling relation, and hence the first priority is to understand galaxies' positions along the curve.  The other significant scalings are then encoded in the ``intrinsic scatter'' about the fundamental curve (discussed and quantified in \S \ref{sec:err} and \S \ref{sec:err2}).

The right panel of Figure \ref{fig:fcurve3d} shows the same data, but now in dMRL space.
Recall that the only difference between dMRL space and MRL space is that the
dynamical mass within the half-light radius,  $M_{1/2}$, is replaced by the 
dark matter mass within the same radius: $M_{1/2} \rightarrow M_{1/2}^{\rm DM}$.
The half-light dark matter mass is determined by subtracting the stellar mass of each
system via $M_{1/2}^{\rm DM} = M_{1/2} - M_*/2$. 
For the E galaxies of \citet{graves09ii}  we use stellar masses derived from the estimates of \citet{gall05mstar} \citep[see][for more details]{graves10iii}.  For the dE sample of \citet{geha03deii}, explicitly computed stellar masses are unavailable so we assign them stellar masses from their observed integrated colors using the prescription of \citet{bell03}.
 For the CSphs and dSphs we assume $M_{1/2} = M_{1/2}^{\rm DM}$, because the dynamical mass-to-light ratios
 in these systems are very large.

The motivation for exploring dMRL space and its fundamental curve is that we
would like to use the dark matter mass density within $r_{1/2}$ as a estimator for the halo virial mass.  
With a virial mass estimate in hand, the fundamental curve relation can be used to 
provide an approximate, average relationship between halo virial mass ($M_{\rm vir}$) and galaxy luminosity ($L$). 
This necessitates comparison to a 1-D dMRL relation, as halo virial masses are a one-parameter family. 
We discuss this effort  in \S \ref{sec:halomatch}.

\subsection{MRL Curve Models}
\label{subsec:mrlcurve}

We have chosen to quantify the fundamental curve by treating $\rhalf$ as the parametric variable.  We fit two relations, one in the $\rhalf - \Lhalf$ plane and another in the $\rhalf - \Mhalf$ plane.  The derived pair of relations (RL and RM) define our fundamental curve relation for the three MRL variables.  We also fit the curve directly in three dimensions for some models, but the derived parameters were effectively identical, and hence we use the simpler two dimensional fits for clarity. We now describe our choice of functional forms for modeling these relations, followed by a set of five best-fit models for the fundamental curve, distinguished by slight differences in the fitting procedure and the choice of $\Mhalf^{\rm DM} = \Mhalf - M_*/2$ as the mass variable in place of the raw $\Mhalf$.

For the $\rhalf - \Lhalf$ relation, we define $\rtil_L \equiv \log (\rhalf/r_L)$ and $\Ltil \equiv \log (\Lhalf/L_0)$ and
employ a fit following the empirically-motivated form
\begin{align} 
\Ltil =  \rtil_L \, \frac{a+b}{2} + \left[ s - \rtil_L (a-b) \right] \frac{\arctan(\rtil_L / w)}{\pi}.
\label{eqn:arctanrvsl}
\end{align}

 Equation \ref{eqn:arctanrvsl} has the property of smoothly transitioning from  an asymptotic slope $a$
 (such that $L_{1/2} \propto r_{1/2}^a$)  for $r_{1/2} \ll r_L$  to  $b$ (i.e. $L_{1/2} \propto r_{1/2}^b$)  for $r_{1/2} \gg r_L$,
 with the width of the transition zone at $r_L$ defined by $w$.  The parameter $L_0$ is then the characteristic luminosity at $r=r_L$, 
 and the final parameter $s$ determines the size of a luminosity offset that occurs in the transition region (e.g. the break in luminosity at $\log(r_{1/2}) \approx 0$ 
 in the upper-middle panel of Figure \ref{fig:fcurve2d}).  This fitting function simply yet generically  captures the behavior of a data 
 set that has distinct asymptotic power laws and a smooth transition region between them.

For the $\rhalf - \Mhalf$ relation we utilize a fitting function with a form identical to Equation \ref{eqn:arctanrvsl}:
\begin{align} 
\Mtil =  \rtil_M \, \frac{\alpha + \beta}{2} + \left[ \sigma - \rtil_M (\alpha - \beta) \right] \frac{\arctan(\rtil_M / \omega)}{\pi}.
\label{eqn:arctanrvsm}
\end{align}
where $\rtil_M \equiv \log (\rhalf/r_M)$, so that $r_M$ defines the transition radius and $\Mtil \equiv \log (\Mhalf/M_0)$ defines a characteristic mass scale $M_0$ at $r=r_M$.  

Using this method, the $M_{1/2}$ vs. $L_{1/2}$ relations are generated by eliminating our chosen parametric variable $r_{1/2}$ in Equations \ref{eqn:arctanrvsl} and \ref{eqn:arctanrvsm}.  For comparison, we also directly fit the ML relation using the form of Equation \ref{eqn:arctanrvsl}, and find very similar relations to those shown below.  Hence, the results presented here are likely not very sensitive to the choice of $r_{1/2}$ as the parametric variable.

Motivated by the fact that we are interested in understanding each type of galaxy universally (CSph, E, dE, and dSph) we weight the data points such that each of the four groups has equal weight (i.e. the weight for each point is $1/N_{type}$ where $N_{type}$ are the number of objects of that type).  Furthermore, for the E data set of \citet{graves09ii}, we weight each point by the relative fraction of galaxies in that particular bin so as to properly represent the full SDSS population rather than the choice of bin locations.  With these weights for the data set, a non-linear least-squared fit for the parameters in Equations \ref{eqn:arctanrvsl} and \ref{eqn:arctanrvsm} (using a Levenberg–Marquardt algorithm) fully determines the one-dimensional relations.  

\begin{figure*}[p!]
\epsscale{1.10}
\plottwo{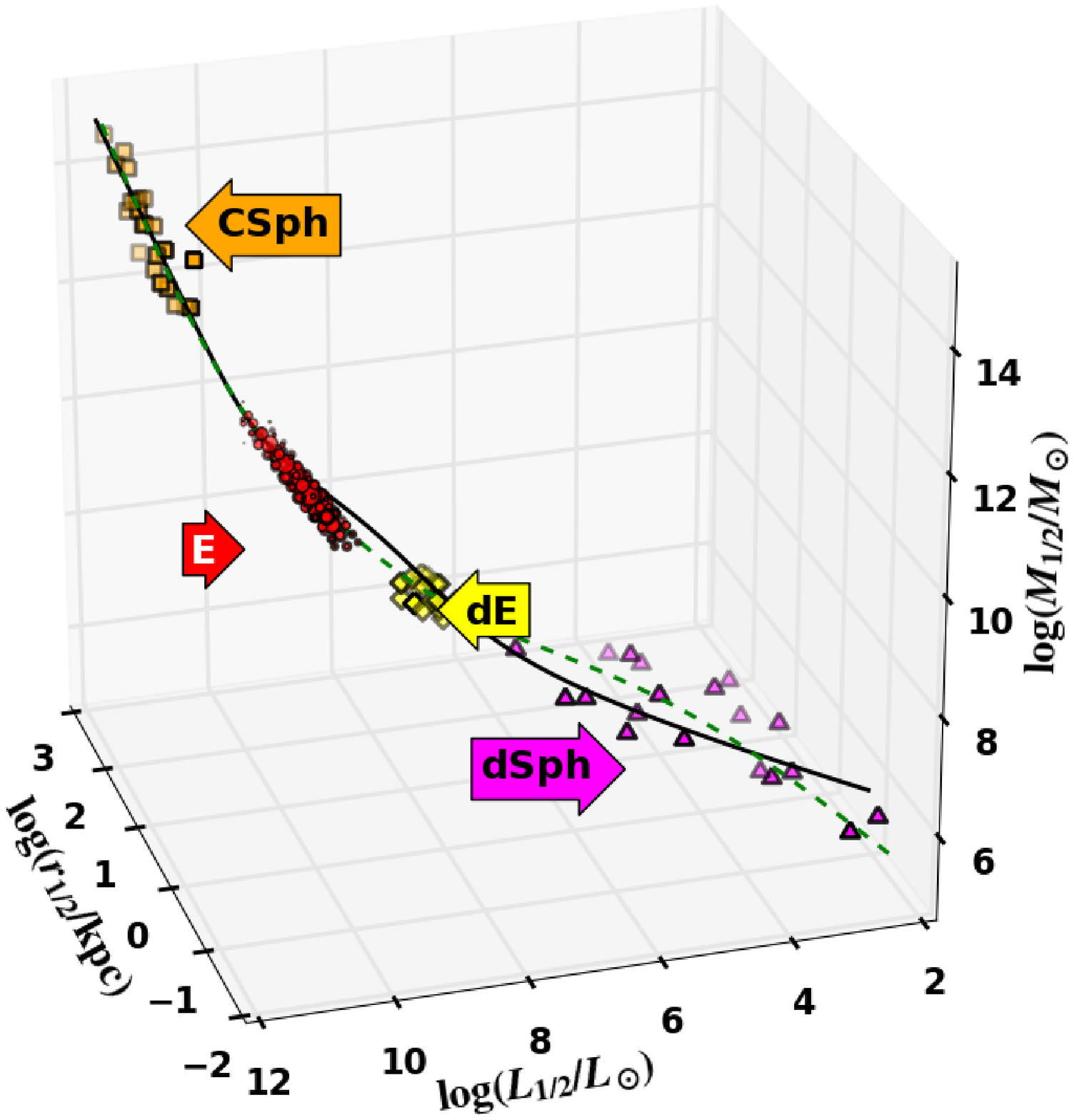}{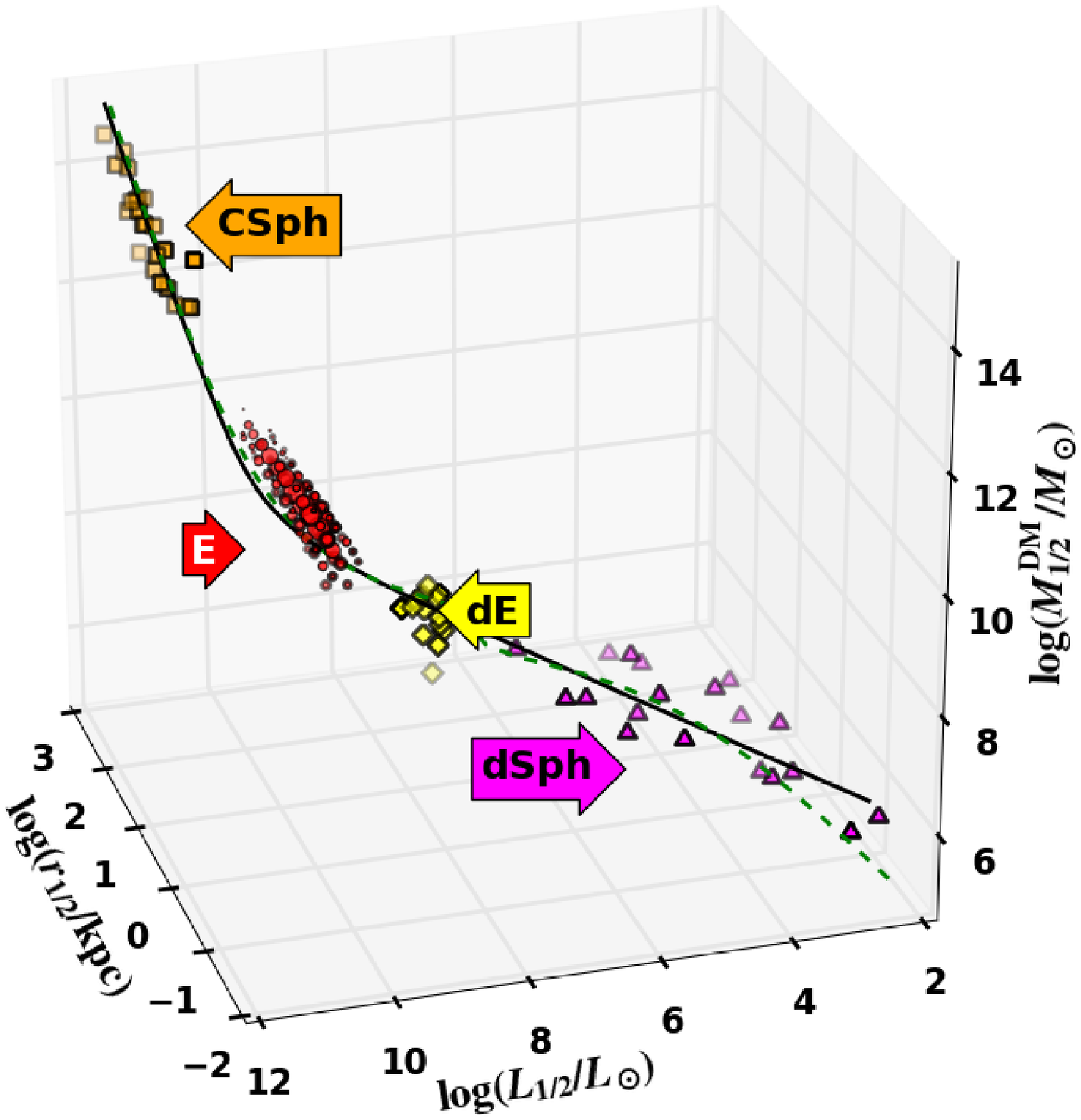}
\caption{Three dimensional representations of the best-fit fundamental curve relations in MRL space (left panel) and dMRL space (right panel, with $\Mhalf \rightarrow M_{1/2}^{\rm DM}$).  Our fiducial models MRL-2 (left panel) and dMRL-2 (right panel) are shown by black solid lines.  Models MRL-1 (left panel) and dMRL-3 (right panel) are shown by a dotted green line. The data point color and point-type scheme matches that of the previous figures (see \S \ref{sec:dat}).  Rotating animations of these plots are available at \movieurl.}
\label{fig:fcurve3d}
\end{figure*}

\begin{figure*}[p!]
\epsscale{1.2}
\plotone{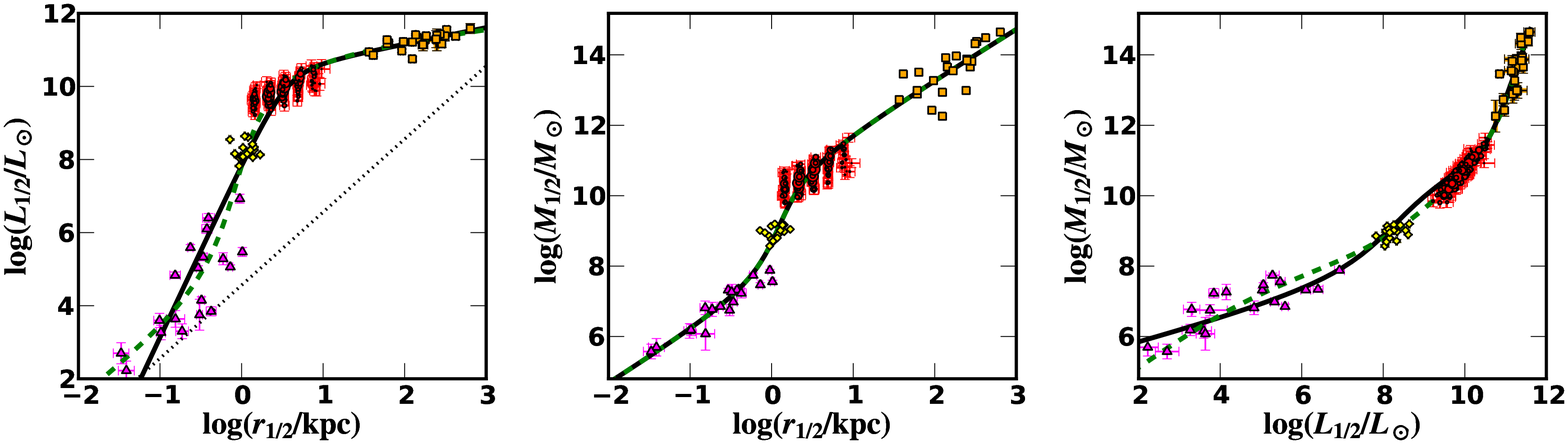}
\plotone{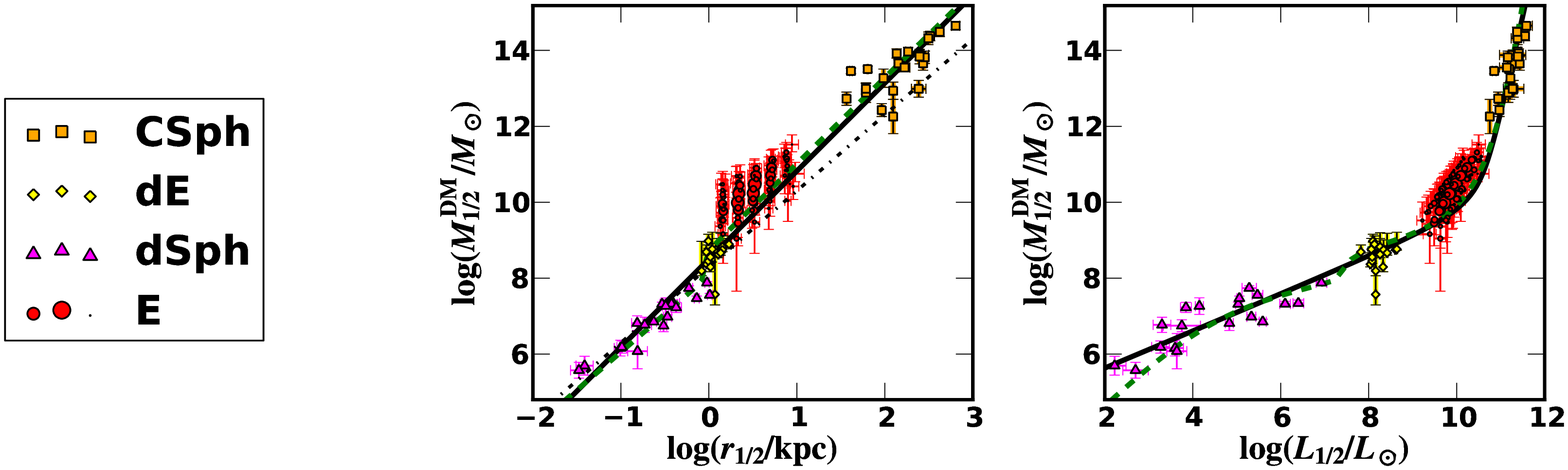}
\caption{Projections of galaxy data and fundamental curve in MRL (upper panels) and dMRL (lower \& upper-left panels) spaces.  The  best-fit fundamental curve models for MRL-1 (upper, green dashed line), MRL-2 (upper, solid black line, fiducial) dMRL-3(lower, green dashed line) and dMRL-2 (lower, solid black line, fiducial) are shown in each projection. The straight dotted line in the left-hand panel represents the detection completeness limit for dSph galaxies found in the SDSS of $\mu_V = 30$ mag arcsec$^{-2}$.  The black dash-dotted line in the lower middle panel is the relation for for dSphs and spiral galaxies from \citet{walker10dsphsp}.    Error bars are observational error bars for each galaxy (see \S \ref{sec:err} for more details).  The data point color and point-type scheme matches that of the previous figures (see \S \ref{sec:dat}).  }
\label{fig:fcurve2d}
\end{figure*}

With the models for the fundamental curve and this fitting procedure, we call our empirically fit fundamental curve model ``MRL-1,'' with best-fit parameters given in the first column of Table \ref{tab:curveparams}.  The relation is shown in projection on the MRL axes as a blue-dashed line on the top panels of Figure \ref{fig:fcurve2d}, along with data points for the individual galaxies and their associated observational error bars (error bars are discussed in detail in \S \ref{sec:err}). We show the same curves and data points as 3-D representations in \ref{fig:fcurve3d}, with the MRL-1 shown as the dashed green line in the left panel.

The black dotted line in the upper left panel of Figure \ref{fig:fcurve2d} shows the surface brightness detection limit for dSphs, $\mu_V = 30$ mag arcsec$^{-2}$ \citep{kop08,walsh09}.   Given that the detection limit indicates that the least luminous dSphs galaxies are at the edge of detectability, it is plausible that the shallow slope in RL at faint $L_{1/2}$ is due to a selection effect. The ``stealth galaxies'' of \citet{bull09stealth}, if present, could substantially alter the slope at the faint end. Thus, we also include an ``MRL-2'' model in which the $s$ parameter is forced to be 0, causing the faint end slope to trace the full dSph population instead of being strongly driven by the faintest of them. This model is shown as the solid black line on the left panel of Figure \ref{fig:fcurve3d} and the upper panels of \ref{fig:fcurve2d}, and the best-fit parameters are given in the second column of Table \ref{tab:curveparams}.  Given the fact that most of the faint dSphs skirt the edge of this detection limit \citep[e.g.][]{walsh09}, we consider the MRL-2 model to be the more robust choice for characterizing the MRL fundamental curve.

The fit parameters listed in Table \ref{tab:curveparams} for the MRL-2 model
reveal that the smallest galaxies with $L \lesssim 2\, L_0 \simeq 4\times 10^9 \, L_\odot$ follow a mass-luminosity
relationship that varies weakly with luminosity
\begin{equation} 
M_{1/2} \propto L_{1/2}^{\alpha/a} \propto L_{1/2}^{0.30} \, ,
\end{equation}
while the largest galaxies ($L \gtrsim 4\times 10^9 \, L_\odot$) 
obey a steep mass-luminosity relationship with
\begin{equation}
M_{1/2} \propto L_{1/2}^{\beta/b} \propto L_{1/2}^{3.2} \, .
\end{equation}
Both regimes are clearly very far from mass-follows-light scalings\footnote{For $r_L$ and $r_M$ values somewhat different from the best-fit for this data set, the values of these slopes can be quite different, but mass-follows-light never holds for any reasonable fits.} (i.e., $M_{1/2} \propto L_{1/2}$).  

For the smallest galaxies, large changes in luminosity correspond to fairly minor changes in half-light mass.
Conversely, for the largest galaxies, a factor of $\sim 2$ change in luminosity corresponds to more than an order of
magnitude change in half-light mass.  This is the same effect noted in \S \ref{sec:mlrspace} (with regard to Figure \ref{fig:mlr_mtol}),
and without any appeal to theory suggests that two qualitatively different processes are acting
to suppress baryon conversion into stars along the transition from small galaxies to large.  The smallest galaxies
seem to be limited by the dark matter mass itself (e.g., by the  potential well depth), 
while the largest galaxies seem to be baryon limited (e.g., by the supply of cool gas for star formation). 

Also of interest is the sharp transition  in the RM relation at $\log(r_{1/2}) \simeq 0.5$ and $\log(M_{1/2}) \simeq 9$, where the half-light mass suddenly jumps with increasing radius.  This transition scale corresponds closely to the point where the dynamical mass-to-light ratios of galaxies reach their minimum (Figure \ref{fig:mlr_mtol}) and thus where baryons contribute substantially to the mass compared to dark matter.  It is possible that this feature is enhanced or even caused by the effects of baryonic contraction \citep{blumenthal86} as discussed in the context of dark matter masses below.

\subsection{dMRL Curve Models}
\label{subsec:dmrlcurve}

Recall that the dMRL relation is distinguished from the MRL relation by the use of $\Mhalf^{\rm DM} = \Mhalf - M_*/2$ as the mass variable in place of the raw dynamical $\Mhalf$. The fit to the data in this space using Equations \ref{eqn:arctanrvsl} and \ref{eqn:arctanrvsm} is our ``dMRL-1'' model.  Trying a variety of starting values for the parameters revealed that $r_M$ is not well-constrained by the data and often would end up outside the data set regardless of the starting value.  Hence, we used the RL relation to set the scale, through the constraint $r_M = r_L$.  Using this constraint, the final parameters are given in the third column of Table \ref{tab:curveparams} and plotted in the right panel of Figure \ref{fig:fcurve3d} and the lower panels of Figure \ref{fig:fcurve2d} as the red dashed line.

As Table \ref{tab:curveparams} shows, the dMRL-1 model best-fit parameters have  $\alpha \approx \beta$, and $\sigma$ preferring 0.  Equation \ref{eqn:arctanrvsm} for dMRL-1 reduces to a power law for $\alpha = \beta$ and $\sigma=0$, so the $\rhalf - \Mhalf^{\rm DM}$ relation turns out to be very close to a single power law (linear in $\Mtil$ and $\rtil_M$).  Hence, the RM relation can be modeled as a simple power law
 \begin{align}
\Mhalf^{\rm DM} = M_0 (r_{1/2}/r_M)^{\alpha},
\label{eqn:powerlawrvsm}
\end{align}
where $r_M$ is determined from the dMRL-1 fit to simplify comparisons.  The value of the slope $\alpha = 2.33$ is also given in Table \ref{tab:curveparams}.  The lower-middle panel of Figure \ref{fig:fcurve2d} compares this fit (red dotted line) to dMRL-1, showing an insignificant difference.

Thus in the second dMRL model (dMRL-2) we adopt Equation \ref{eqn:powerlawrvsm} as the model for the RM relation, and the RL model of MRL-2, selected due to the likely presence of the stealth galaxy selection effect. We tabulate the best-fit parameters for this model in the second-to-last column of Table \ref{tab:curveparams}, and plot it as the black solid line in the lower panels of Figure \ref{fig:fcurve2d} and the right panel of Figure \ref{fig:fcurve3d}.  

In the RM relation  of the dMRL space, we include for comparison the \citet{walker10dsphsp} relation derived using Milky Way dSphs for the faint end and spiral galaxy rotation curves for the galaxy regime (black dash-dotted line on the lower-middle panel of Figure \ref{fig:fcurve2d}).  We note here that while the \citet{walker10dsphsp} non-dSph sample is a very different set of galaxies that may obey different scaling relations from our sample\footnote{See \citet{mg10} for a discussion of how dSph scaling relations connect to spirals.}, it is fairly close to our relation in the galaxy regime.  However, the relation steepens with the inclusion of Es and CSphs, so our derived slope is somewhat higher than a $M^{\rm DM} \propto r^2$ relation.

Motivated partly by this $M^{\rm DM} \propto r^2$ result on the faint end, as well as the greater uncertainty in $M_{1/2}^{\rm DM}$ for the dEs and Es (see \S \ref{sec:err} and \ref{sec:err2}), we consider a third dMRL model (dMRL-3).  In this model we use Equation \ref{eqn:arctanrvsm} for the RM relation, but we force the faint-end slope to 2 and set the normalization to pass through the dSphs.  We then force the $r_M$ scale to match $r_L$ (from MRL-2), set $\omega=0.01$  to ensure a small transition rgion, and fir the remaining parameters.  We also continue to use the RL model of MRL-2 for dMRL-3.  In the last column of Table \ref{tab:curveparams}, we show the best-fit parameters of this model, and in the lower panels of Figure \ref{fig:fcurve2d} and the right panel of Figure \ref{fig:fcurve3d}, we plot it as the green dashed line.

Before continuing, we note a discrepancy for the E galaxies in the dMRL models, most apparent in the lower-middle panel of Figure \ref{fig:fcurve2d} -- the Es tend to have higher $\Mhalf^{\rm DM}$ than the best-fit relations.  Recall, however, that the primary motivation for exploring the $\Mhalf^{\rm DM}$ as a parameter is that it will allow us to map galaxy properties to an underlying dark matter halo mass.  This mapping is hindered somewhat by the contraction of baryons.  An anomalously high dark matter mass for the galaxies with the highest baryonic-to-dark matter ratio is precisely what is expected if dark matter halos contract due to central condensation of baryonic matter \citep{blumenthal86}. Thus, we might expect an offset in the scaling relations of galaxies at the scale where baryonic condensation has been the most significant.  In  \S \ref{sec:halomatch} we estimate the degree to which baryonic contraction has increased the $\Mhalf^{\rm DM}$ masses in our E galaxy sample and show that this increase approximately accounts for the discrepancy.   Further, as discussed more in \S \ref{sec:err}, a power law for the RM relation is in general more robust to the problem of a non-monotonic mapping of baryonic galaxies to dark matter halos. Thus, use of a power law for the RM model is a reasonable choice for the exercise of halo profile matching (described in \S \ref{sec:halomatch}), while still being a decent fit to this data set. In the RL space, as described above for MRL-2, it is more appropriate to use the $s=0$ model so as to prevent the stealth galaxies selection effect from strongly biasing the faint end slope. Thus, we adopt dMRL-2 as our fiducial model in the latter sections of this paper.

\begin{deluxetable}{|cccccc|}
\tablecolumns{6}
\tablecaption{Fundamental Curve Model Parameter Values for Equations \ref{eqn:arctanrvsl} and \ref{eqn:arctanrvsm}.}
\tablehead{
\colhead{Model Name} &
\colhead{MRL-1} &
\colhead{MRL-2\tablenotemark{+}} &
\colhead{dMRL-1} &
\colhead{dMRL-2\tablenotemark{*}} &
\colhead{dMRL-3}
}

\startdata

Mass Variable & $M_{1/2}$ & $M_{1/2}$ & $M_{1/2}^{\rm DM}$ & $M_{1/2}^{\rm DM}$ & $M_{1/2}^{\rm DM}$ \\
\hline
RM Model & Eqn. \ref{eqn:arctanrvsm} & Eqn. \ref{eqn:arctanrvsm} & Eqn. \ref{eqn:arctanrvsm} & Eqn. \ref{eqn:powerlawrvsm}  & Eqn. \ref{eqn:powerlawrvsm} \\
\hline
$\log(r_L/{\rm kpc})$ & -0.04 & 0.54 & -0.04 & -0.04 & -0.04 \\
$\log(L_0/L_\odot)$ &  7.54 & 9.95 &  7.54 &  7.54 &  7.54 \\
$a$ & 1.67 & 4.77 & 1.66 & 1.66 & 1.66 \\
$b$ & 0.26 & 0.44 & 0.26 & 0.26 & 0.26 \\
$w$ & 0.32 & 0.42 & 0.32 & 0.32  & 0.32  \\
$s$ & 6.58 & 0 & 6.58 & 6.58  & 6.58  \\
\hline 
$\log(r_M/{\rm kpc})$ & 0.09 & 0.09 & -0.04 & \nodata & -0.04 \\
$\log(M_0/M_\odot)$ & 9.12  & 9.12 & 8.40 & 8.50 & 8.32 \\
$\alpha$ & 1.44 & 1.44 & 2.33 & 2.32 & 2.00\\
$\beta$ & 1.42 & 1.42 & 2.28 & \nodata & 2.27\\
$\omega$ & 0.27 & 0.27 & 0 & \nodata & 0.01 \\
$\sigma$ & 3.13 & 3.13 & 0 & \nodata & 0.69

\enddata

\tablenotetext{+}{Fiducial MRL Model}
\tablenotetext{*}{Fiducial dMRL Model}

\label{tab:curveparams}
\end{deluxetable}

\subsection{Scatter and Uncertainty in the Fundamental Curve}
\label{sec:err}

It is interesting to ask about the degree of intrinsic scatter within the fundamental curve that was defined in the
previous section, but in order to do that we need to estimate the observational uncertainties on the MRL variables.
Representative error bars for $M_{1/2}$, $r_{1/2}$, and $L_{1/2}$ are shown in Figure \ref{fig:mlr_mtol} 
 for several galaxy types.  Observational errors for $M_{1/2}^{\rm DM}$ are presented in Figure \ref{fig:fcurve2d}.  Individual error bars for each data point are shown in Figure \ref{fig:fcurve2d}.
 Note that for the  faint dSphs and the CSph, the measured mass-to-light ratios are much larger than any reasonable stellar population (e.g. $M_{1/2}/L_{1/2} >> 1$).  Hence, they are dark matter-dominated ($M_{1/2} \approx M_{1/2}^{\rm DM}$), and hence the $M_{1/2}$ and $M_{1/2}^{\rm DM}$ errors are similar to each other. 
 For the dE and E galaxies, however, the mass-to-light ratios are closer to that expected of stellar populations and hence a significant amount of mass within $r_{1/2}$ is in stars rather than dark matter, so $M_{1/2}^{\rm DM}$ errors are larger for these objects due to the errors on $M_*$.  
 
For the E galaxies, the uncertainty in $M_*$ due to stellar populations is a major uncertainty.  While the observational errors play a role in general, for the large stacked E data sets here, the errors are certainly dominated by systematics, of which there are three major components \citep{graves10iii}.  First, there is variation due to the method used to derive $M_*/L$ (e.g. integrated colors or particular spectral features).  As shown in \citet{graves10iii}, this contributes a $1\sigma$ scatter of $\sim 0.08$ dex.  Second, the assumed star formation history affects the inferred stellar mass, at the level of 0.15 dex for this sample \citep{graves10iii}.  Third, the choice of IMF has a major effect on the inferred $M_*$.  For the example (conservative) comparison of Chabrier as compared to Kroupa \citep{long09}, the inferred $M_*$ varies by 0.26 dex.   
More detailed studies of individual objects can potentially reduce the systematics \citep[e.g.][]{cappellari06}, but the analysis above is appropriate for the large data set in use here.
Thus we show error bars by adding the above 3 components in quadrature, providing a factor of 2 uncertainty in the $M_*$ used for mapping $M_{1/2}$ to $M_{1/2}^{\rm DM}$.  
This error on the Es is shown in Figure \ref{fig:fcurve2d} as the error bar on $M_{1/2}^{\rm DM}$, and we also adopt it in the next sections as the error for $M_{1/2}^{\rm DM}$.  

The error bars shown in Figure \ref{fig:fcurve2d} account for the uncertainty in measuring the dark matter mass as it is today, but do not include the systematic uncertainty that remains in our ability to map an observed dark matter density to the virial properties of that dark matter halo.  
Baryonic contraction \citep{blumenthal86} in particular can make the mapping between density and global halo mass quite difficult. 
We expect this uncertainty to be particularly important for E galaxies because they have the highest baryon fractions.   We discuss this effect in more detail in \S \ref{sec:halomatch}.

For the dE sample, $M_*$ is inferred from SDSS colors as described in \S \ref{sec:halomatch}. Errors can be estimated from this procedure by comparing the inferred $M_*$ for each band.  Using this procedure, the scatter in the inferred $M_*$ is about $30\%$, comparable to the observational errors for $\Mhalf$.  This estimate has its own set of systematic errors like those described above -- we do not quantify this here due to the smaller sample size (and hence larger random errors) and more simplistic method compared to the Es.  Regardless, the error bars are large enough to be consistent with the fundamental curve.

For the CSph population, the uncertainty in $\Mhalf$ is difficult to characterize, as it is primarily due to the use of the galaxies to trace the velocity dispersion instead of the ICS.  The effect this will have is not as well understood, as represented by the ``?'' in the error bar of Figure \ref{fig:mlr_mtol}.  The simulations of \citet{dolag10csphsims} find a disagreement in $\sigma$ of $\sim 20\%$ between galaxies and the ICS component (i.e. approximately 50\% in mass), although this is not necessarily representative of the clusters in our sample.  In order to broadly represent this uncertainty, we have assumed a factor of 2 uncertainty on $\sigma$ in deriving the error bars in the next section.

Adopting these observational error bars, it is clear from the top panel of Figure \ref{fig:fcurve2d} that the actual scatter about the fundamental curve is larger than the observational errors. We estimate the scatter by  computing the residuals of $\Mhalf$ and $\Lhalf$ from the fundamental curve, and measure the standard deviation with weights as described in \S \ref{sec:curve}. The resulting as-observed scatter in $M_{1/2}$ at
fixed $r_{1/2}$ about the MRL-2 relation is $\delta \log \Mhalf = 0.41$.  Subtracting  the observational error in $\Mhalf$ (including the contribution due to the error in $\rhalf$) in quadrature from this value, we obtain an estimated intrinsic scatter of $\Delta \log \Mhalf = 0.38$.    
Using the dMRL-2 relation, the observed scatter in $M_{1/2}^{\rm DM}$ at fixed $r_{1/2}$ is $\delta \log M_{1/2}^{\rm DM} = 0.60$ and the intrinsic is $\Delta \log M_{1/2}^{\rm DM} = 0.20$, due to the larger uncertainties in $\Mhalf^{\rm DM}$. We emphasize that this estimate of intrinsic scatter is only approximate, given our small samples size and our rough characterization of observational errors over the entire (disjoint) population of our objects.

Applying the same method to the RL relation (identical for MRL-2 and dMRL-2), we get an observed scatter in $\Lhalf$ at fixed $\rhalf$ of $\delta \log \Lhalf = 0.73$, and estimated intrinsic scatter $\Delta \log \Lhalf = 0.71$.  This is relatively high, but is driven almost entirely by a few dSph outliers (the low dSph points in the upper-left panel of Figure \ref{fig:fcurve2d}) that render the distribution non-Gaussian.  The dSphs generally have relatively high error bars, but this is not accounted for in the averaging process above.  Thus, removing the discrepant dSphs gives an observed scatter of  $\Lhalf$ at fixed $\rhalf$ of $\delta \log \Lhalf = 0.42$, and $\Delta \log \Lhalf = 0.37$.

These values for the scatter are purely empirical measurements of the deviation of individual galaxies from the fundamental curve.  As discussed in \S \ref{sec:curve}, the intrinsic portion of this scatter encodes all of the additional scalings in galaxy formation that are sub-dominant to the curve itself.  In the next section, we describe theoretically expected scatter based on the profile matching scheme.

\section{Dark Matter Profile Matching}
\label{sec:halomatch}

We now describe a technique to use the fundamental curve described in the last section to derive global relations connecting dark matter halos to the luminous properties of the galaxy.  The main relationship we would like to derive is the median
relation between $M_{\rm vir}$ and $L$.  We refer to this method as ``profile matching,'' as it matches the mass profile of galaxies to dark matter halos to do this.   While the analysis presented here relies on NFW halos \citep*{NFW97} in $\Lambda$CDM, the general approach  is applicable to any halo form or variant cosmology.

\begin{figure*}[h!]
\plotone{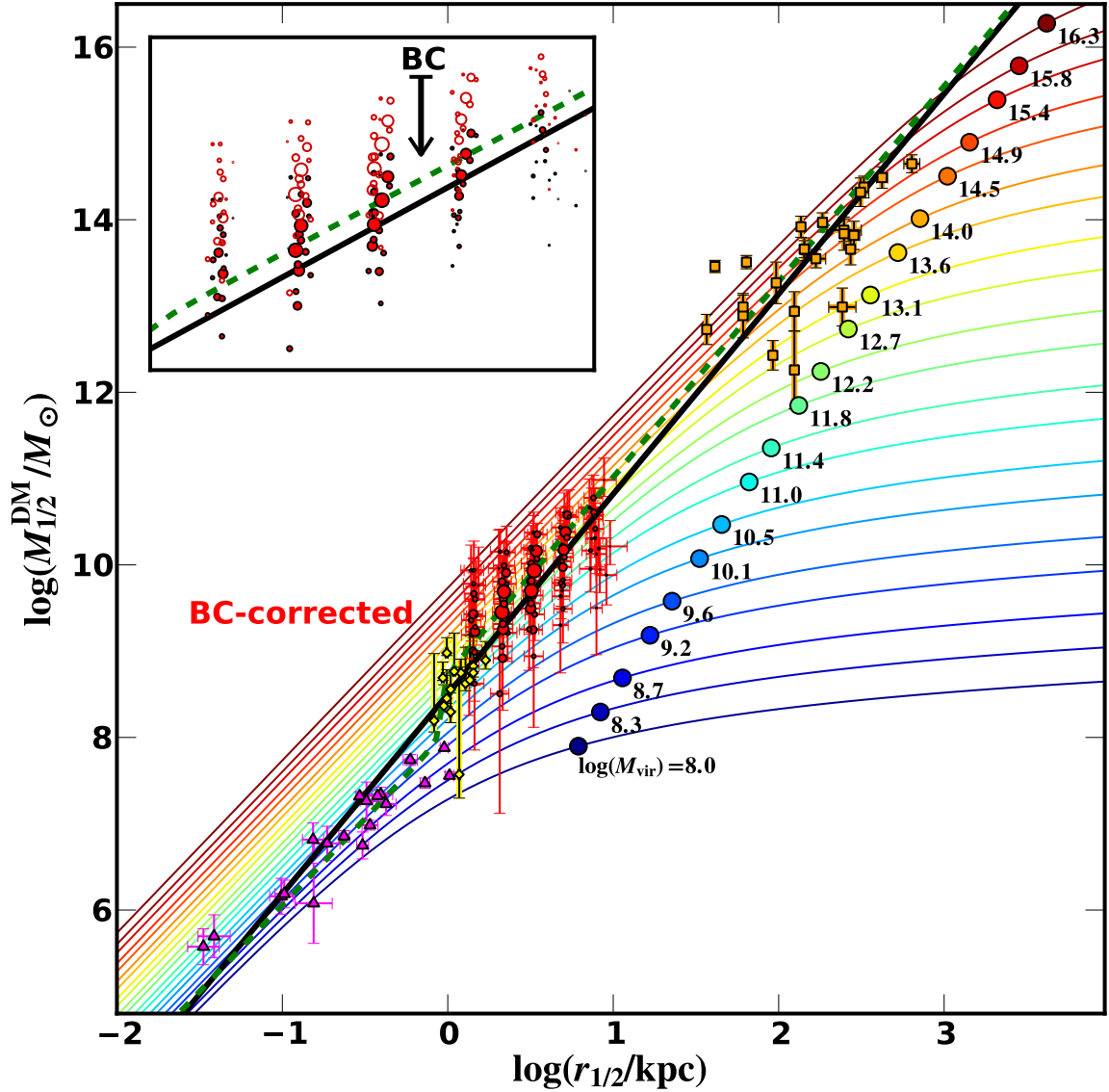}
\caption{Profile matching schematic.   The solid colored lines are the enclosed mass as a function of distance from the center for NFW profiles of a given mass assuming a $c$-$M$ relation as described in the text.  Their virial radii are enumerated at $R^{\rm halo}_{1/2}$ (i.e. $M(R^{\rm halo}_{1/2})=M_{\rm vir}/2$) as large colored points.  The black solid line is our fiducial fundamental curve relation (dMRL-3) projected into this space, while the green dashed line is dMRL-3 (see \S \ref{sec:curve}).  
The points are the data points used to fit the fundamental curve (see \S \ref{sec:dat} for meaning of colors and symbols).  Observational error bars are shown for $\Mhalf^{\rm DM}$ and $r_{1/2}$, derived as described in \S \ref{sec:err}.  We note that the observational error bars on each point are significant and this contributes to the apparent scatter in the data. 
For the E data points, we correct the observed $M_{1/2}^{\rm DM}$ for baryonic contraction using the \citet{blumenthal86} adiabatic contraction estimate described in the text, which results in an offset of $\sim .5$ dex in $\log(M_{1/2}^{\rm DM})$.
In the inset, we show the E galaxies before this correction as red open circles, and the solid red points with black outlines are the corrected values. }
\label{fig:massplot}
\end{figure*}

$\Lambda$CDM simulations predict that at a fixed physical radius $r$, a more massive dark matter halo will be denser, on average, than
a less massive dark matter halo \citep[e.g.][]{NFW97}.  Moreover, the {\em typical} mass
profile for a given virial mass halo is determined by the virial mass in a one-to-one way, such that knowledge
of $M_{1/2}^{\rm DM}$ and $r_{1/2}$ for a  galaxy can be mapped to the unique dark matter halo virial mass that gives $M_{\rm halo}(r=r_{1/2};M_{\rm vir}) = M_{1/2}^{\rm DM}$. 
 Of course, this mapping is not without scatter, and we address this issue in \S \ref{sec:err2}. This mapping is also made more difficult by the fact that some of the galaxies we consider reside within subhalos.  We also address this point in \S \ref{sec:err2}.

We assume that each galaxy resides at the center of a dark matter halo and that galaxies have $M_{1/2}^{\rm DM}$, $r_{1/2}$, and $L_{1/2}$ values specified by the dMRL fundamental curve.  We also assume that the dark matter densities within $r_{1/2}$ can be mapped to a virial mass using density scaling relations derived for dark matter halos from {\em dissipationless} simulations.  This is a reasonable assumption for most of our galaxies because most of them are dark matter dominated.  This is not a good assumption for  E galaxies, which have fairly high baryon mass fractions and have likely had their dark matter masses enhanced within $r_{1/2}$ by baryonic contraction \citep{blumenthal86,gnedin04,napo10cendm} .   But as discussed in the previous section, the dMRL curves tend to lie below the dark matter masses in E galaxies in dMRL space.  Indeed, we will show that a first-order correction for the effects of baryonic contraction yields ``uncontracted'' masses for E galaxies that sit along our dMRL fits.

We consider an ensemble of dark matter halos with a range of virial masses $ 10^7 < M_{\rm vir}/M_\odot < 10^{16.5} $.   Each halo is assumed to follow an NFW mass profile $M_{\rm halo}(r) = M(r; M_{\rm vir})$
with a concentration parameter ($c \equiv r_{\rm vir}/r_s$) set by the
{\em median}  concentration-mass relations provided by \citet{klypin10bolshoi} from the Bolshoi simulations.
This simulation
was run with cosmological parameters ($n=0.95$, $\sigma_8 = 0.82$, $h=0.7$, and $\Omega_m = 0.27$) that are very similar to those favored by WMAP7 \citep{WMAP7}.  
We define virial mass and virial radius as in \citet{klypin10bolshoi}, using the virial overdensity
as calculated by the spherical collapse approximation.  
Note that we have extrapolated their fitted concentration-mass-redshift relation to masses beyond those directly probed by the Bolshoi simulation ($M_{\rm vir} = 10^{10.3-14.5} M_\odot$).  However, these extrapolations are  consistent with the scaling behaviors expected from previous simulations that have probed higher and lower mass regimes directly \citep[e.g.][]{neto07,springel08aquarius,mac08}.

The implied dark matter mass profiles for many different virial masses are illustrated as colored lines in Figure \ref{fig:massplot}.  For reference, the half-mass radii for the {\rm dark matter halos}, $R_{1/2}^{\rm halo}$, are plotted
as large colored circles at their associated half-mass values, $M_{1/2}^{\rm halo} = M_{\rm vir}/2$.  The slope of this
relation is almost exactly $M_{1/2}^{\rm halo} \propto (R_{1/2}^{\rm halo})^3$, and therefore significantly steeper than
the $M_{1/2}^{\rm DM} \propto r_{1/2}^{2.3}$ slope favored by our fiducial fit to the fundamental curve of stellar systems.  
The virial mass associated with each mass profile plotted
is indicated to the right of the associated colored circle.

\begin{figure}[htbp!]
\epsscale{1.2}
\plotone{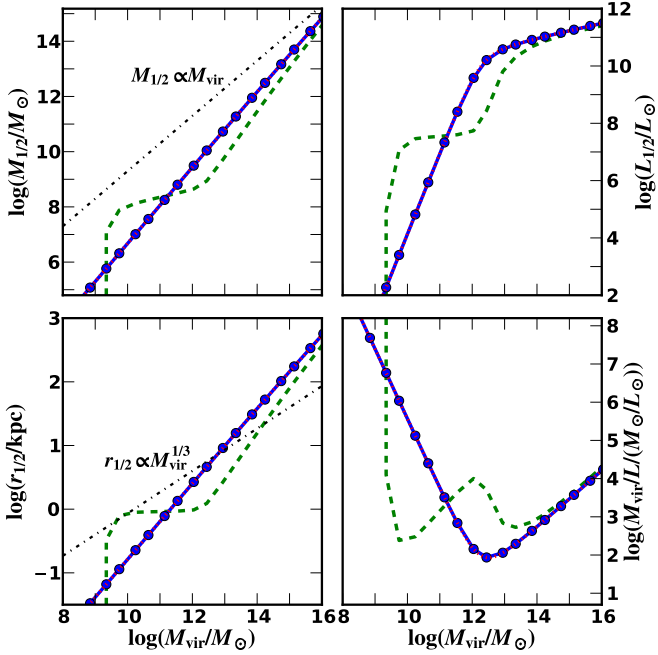}
\caption{Global properties of dispersion-supported galaxies embedded in dark matter halos.  All panels have virial mass on the horizontal axes.  The vertical axes are half-light mass (upper-left), half-luminosity (upper-right), 3-D half-light radius (lower-left), and \emph{virial} mass-to-light ratio. The blue thick lines with points are the inferred relations for the dMRL-2 model, and the dotted red lines are the associated fits using Equation \ref{eqn:matchfit} and Table \ref{tab:matchtab}.  The thick green dashed line corresponds to the dMRL-3 model.  The black dash-dotted line in the upper-left panel shows $M_{1/2} = f_{\rm baryon} M_{\rm vir}$.  The black dash-dotted line in the lower-left panel shows $r_{1/2} \propto M_{\rm vir}^{1/3}$, which is the scaling one might naively expect in the case that stellar radii scale self-similarly with virial radii $r_{1/2} \propto R_{\rm vir}$. 
}
\label{fig:matchplots}
\end{figure}

Overlaid on Figure \ref{fig:massplot} as a thick, black solid line is the $M_{1/2}^{\rm DM}$ vs. $r_{1/2}$ relation for our preferred fundamental curve fit (Model dMRL-2 in Table  \ref{tab:curveparams}). 
The thick green, dashed line is the alternative dMRL-3 relation.
  Each point along these curves is mapped to a single luminosity via its respective dMRL relation. Each
point on the line can also be mapped in a one-to-one way to a median dark matter halo virial mass, set by the particular
$M_{\rm halo}(r) = M(r; M_{\rm vir})$ halo line it intersects. This allows us to back out an implied median relationship between galaxy luminosity and halo virial masses across the range of galaxies considered. Figure \ref{fig:matchplots}  shows the implied $M_{\rm vir}-L$ mapping for each of these curves (dMRL-2, solid blue with points and dMRL-3, green dashed) in the upper right panel.
 Associated relationships between $M_{\rm vir}$ and the other fundamental curve parameters are shown in the other panels of Figure \ref{fig:matchplots}.  Full analytic descriptions of these relation are provided in Appendix \ref{apx:matchfit} (see Table \ref{tab:matchtab}).  For dMRL-2, the  $M_{1/2}$ vs. $M_{\rm vir}$ and $r_{1/2}$ vs, $M_{\rm vir}$ relations are fairly well characterized by power-laws with  $\Mhalf \propto M_{\rm vir}^{1.36}$ and $\rhalf \propto M_{\rm vir}^{1.59}$.  The $L$-to-$M_{\rm vir}$ relation, meanwhile, can be approximated on the faint end as $L \propto M_{\rm vir}^{2.84}$ and on the bright end as $L \propto M_{\rm vir}^{0.26}$.  As expected from our $M_{1/2} - L_{1/2}$ scalings, one interpretation is that mass is the limiting factor in galaxy formation for faint galaxies while baryonic feedback of some kind limits galaxy formation for bright galaxies.

Returning to  Figure \ref{fig:massplot}, we have also plotted the galaxy data points used to fit the fundamental curve
as colored symbols, with error bars reproduced from the lower middle panel of Figure \ref{fig:mlr2d}. The symbol types are the same as those described in \S \ref{sec:dat} and Figures \ref{fig:mlr_mtol}-\ref{fig:fcurve2d} except for the red (E) points, as described below.
Clearly, these points exhibit a large scatter at fixed radius.  As we discuss (and illustrate) in the
next section, one of the reasons for the apparent scatter and offsets is that the measurement errors on each data point are quite large.  
This is particularly important for the red symbols (Es), for which small errors in stellar mass estimation can propagate to very large errors 
in the dark matter masses plotted, potentially in a systematic way.  We discuss inherent vs observational scatter in detail in \S \ref{sec:err2}.

Another effect that adds uncertainty to the mapping between halo mass and galaxy luminosity is baryonic contraction \citep{blumenthal86,gnedin04}, which  increases the dark matter density within a given radius from what it otherwise would have been absent the infall of baryons. The E points (red circles) in Figure \ref{fig:massplot} have been modified  in their $M_{1/2}^{\rm DM}$ masses from those shown in Figures \ref{fig:fcurve3d} and \ref{fig:fcurve2d} in order to approximately account for this effect.  Specifically, the DM masses for the E galaxies in this plot are estimates of the ``intrinsic'' dark matter masses within $r_{1/2}$ prior to the infall of baryons.  We make this estimate  using the \textit{contra} code of \citet{gnedin04} applied to the E galaxy bin with the largest number of galaxies.

In order estimate the degree of the mass enhancement from baryonic contraction, we assume that the initial virial mass followed is that implied by our fiducial curve in Figure \ref{fig:matchplots} (dMRL-2) for the $r_{1/2}$ of the chosen E bin.  We use the concentration-mass relation discussed above to determine the $c_{\rm vir}$ for an NFW profile.  For simplicity we assume a \citet{hernq90} model for the stellar distribution with $M_*$ and $r_{1/2}$ set by the E bin. We determine the ratio of the mass within $r_{1/2}$ before and after the contraction, and correct our profile matching $M_{1/2}^{\rm DM}$ by this ratio.  The points shown in Figure \ref{fig:massplot} assume the
\citet{blumenthal86} adiabatic contraction formula, but we find that with both the \citet{gnedin04} and \citet{blumenthal86} methods, the correction  is large enough to move the E galaxies onto the dMRL-2 relation. 
For simplicity, the error bars on the E points here are simply scaled versions of the direct uncertainty in $M_{1/2}^{\rm DM}$ as presented in Figure \ref{fig:fcurve2d} and do not include the additional uncertainty in the baryonic contraction correction, which is certainly large but hard to quantify. The errors shown here are conservatively small for this reason.

The uncertainty in profile matching in the E/dE regime is nicely illustrated by the differences between the solid curve (from dMRL-2) and
green dashed curves (from dMRL-3) in Figure \ref{fig:matchplots}.   The dMRL-3 relation yields bumps (e.g. a plateau in $L$ around $M_{\rm vir} \sim 10^{12} M_{\odot}$ ) due to the enhanced $M_{1/2}^{\rm DM}$ at  $\log(r_{1/2}) \sim 0$ associated with this relation.  This break in the MR relation maps onto an increased $M_{\rm vir}$, creating this unexpected
feature, which is likely an artifact of baryonic contraction, possibly with a component due to uncertainties in $M_*$.

Regardless of the nature of this bump, however, this dMRL-3 scaling does a slightly better job in matching the properties of the faintest galaxies, as it was designed to have an MR relation that is overweighted in dSph regime (compare the dashed and solid lines in Figure \ref{fig:massplot}). Interestingly, the green dashed curves in Figure \ref{fig:matchplots} reveal features in the scaling relations of the smallest galaxies at $M_{\rm vir} \sim 10^9 M_\odot$ in the form of a wall in $M_{\rm vir}$. Strictly speaking, this is a breakdown in monotonicity of the $L-M_{\rm vir}$ relation (discussed further in \S \ref{subsec:abund}), but for dMRL-3 this is because $M_{\rm vir}$ is very nearly constant with $L$. This might be indicative of a common mass scale for small galaxies   \citep{stri08nat,pen08scale,ok09scale,wolf09} under which luminous galaxies do not inhabit dark matter halos. Abundance matching does not constrain the existence of such a scale, as the galaxies in those halos are too faint to be observed in statistically significant quantities outside the Local Group.  As we discuss below, profile matching is just approaching the point where we can begin to test this possibility as part of a global relation.

\subsection{Comparison to Abundance Matching}
\label{subsec:abund}

\begin{figure*}[htbp!]
\epsscale{1.15}
\plottwo{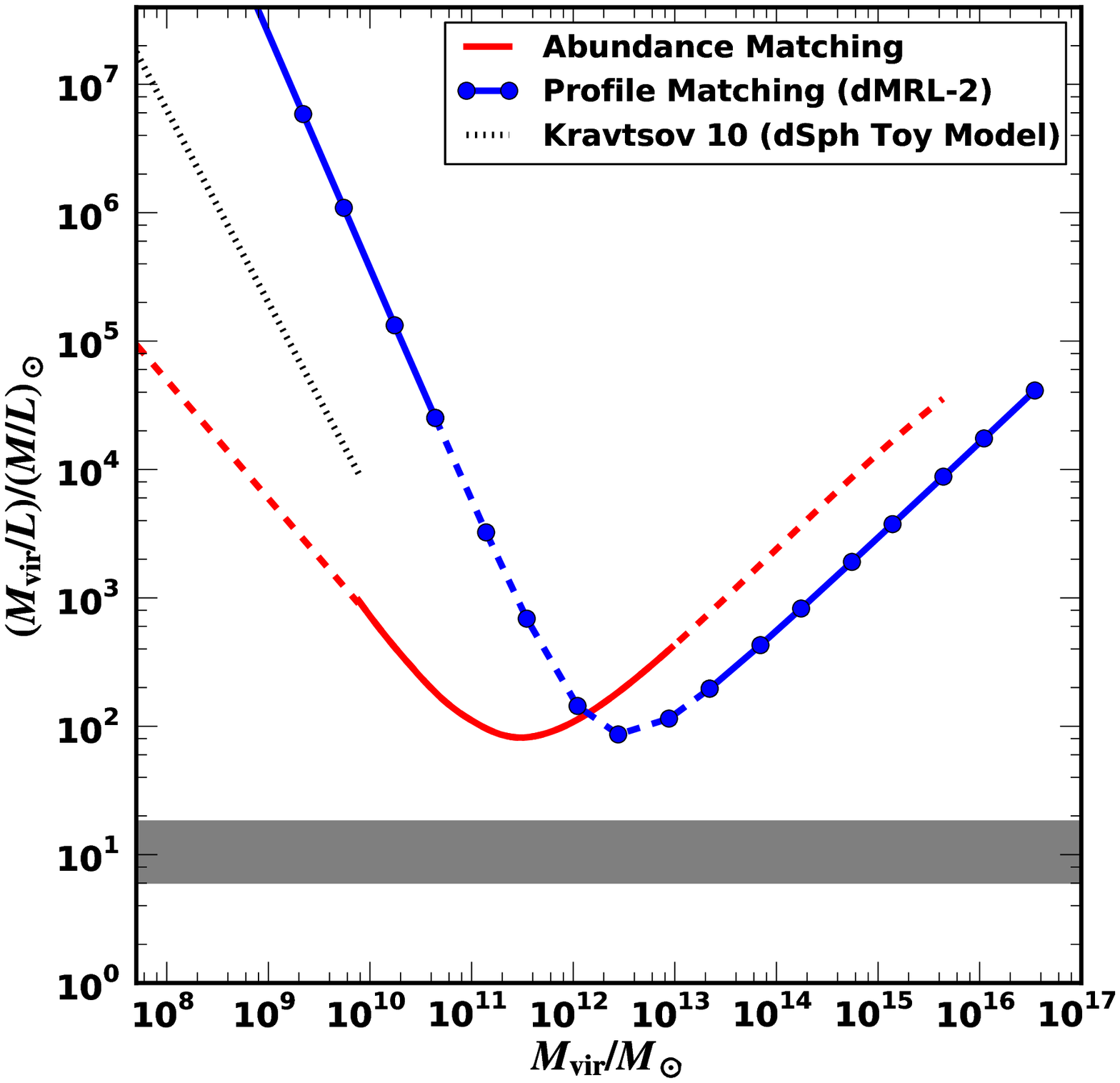}{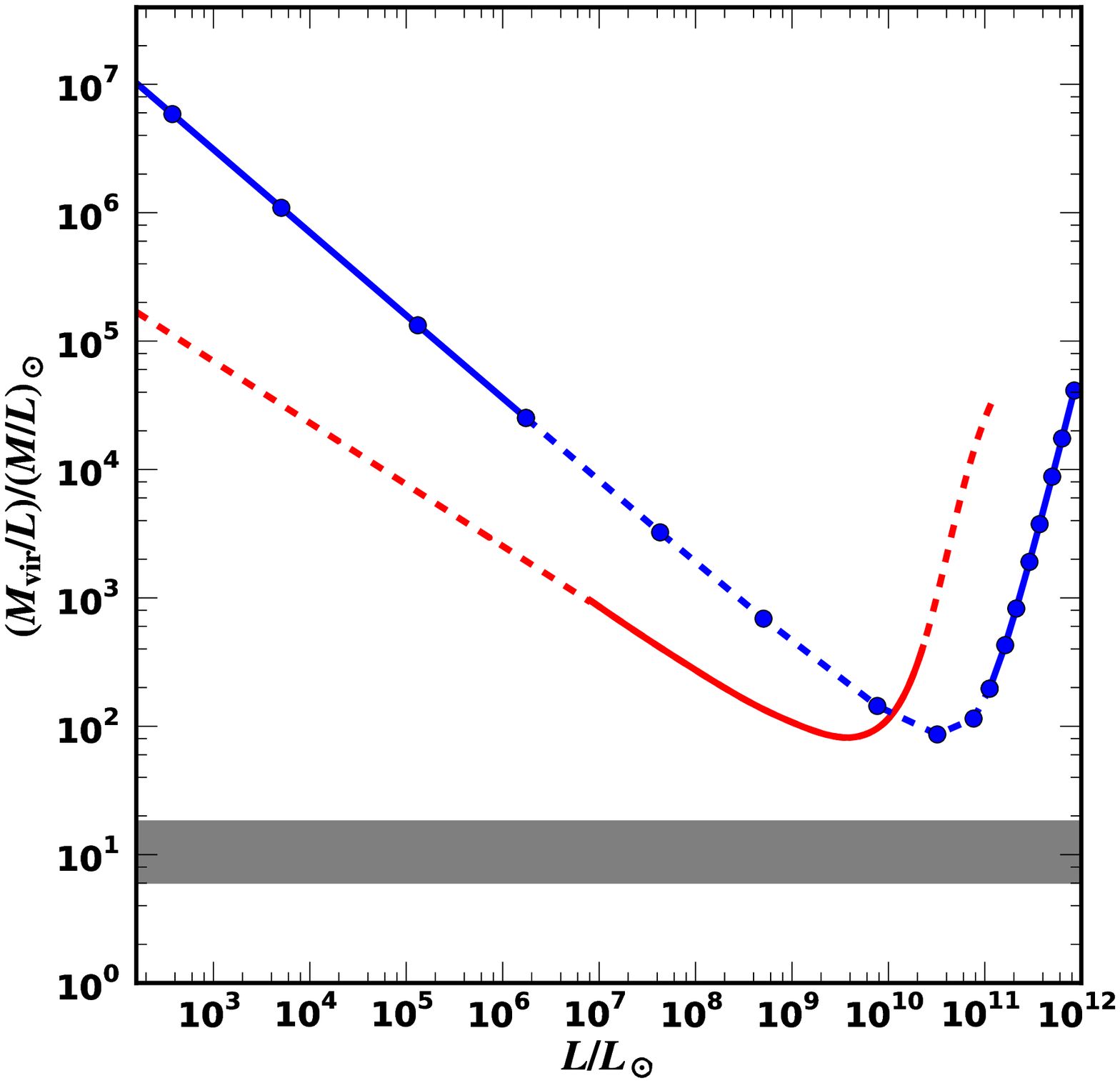}
\caption{Comparison between profile matching and abundance matching. The plots show averaged relations for virial mass-to-light-ratio against virial mass (upper panel) and luminosity (lower panel).  The (blue) line with points is the relation inferred from profile matching as described in this work for the fiducial dMRL-2 model, with the dashed region corresponding to the region for which baryonic contamination of $M_{\rm vir}$ is important.  The (red) solid line without points is from abundance matching the luminosity function of \citet{blanton05dwarfs} to the mass function of \citet{tinker08}, dashed on the bright end where the luminosity function has significant errors due to small numbers.  The (red) dashed line on the faint end is a power law extrapolation of the faint end of the abundance matching relation.  The (black) dotted line is the faint end extrapolation suggested in \citet{krav09drev} to match the luminosity function of MW dSphs (see \S \ref{sec:err2}).  The gray shaded region is the expected mass-to-light-ratio for a system in which all of its baryons are converted into stars, computed as $\Upsilon_*/f_{\rm baryon}$, using a range of $1 < \Upsilon_* < 3$.  }
\label{fig:abundmatch}
\end{figure*}

Figure  \ref{fig:abundmatch} compares our fiducial profile matched results (blue lines, dMRL-2) to those of the independent technique of abundance matching (red lines). The implied ratios $(M_{\rm vir} - L)$ vs. $M_{\rm vir}$ are shown in the left panel and the equivalent relations for $(M_{\rm vir} - L)$ vs. $L$ are shown in the right panel. The blue profiled-matching lines are shown as dashed in the regime where the average dynamical mass-to-light ratio within $r_{1/2}$ is indicative of a significant stellar component, with $M_{1/2}/L_{1/2} < 9$. The line is solid in the regime where our stellar mass subtraction is less important for the dark matter mass determination within $r_{1/2}$. The line types emphasize the point that our profile matching technique is most trustworthy in the luminosity/mass extremes. We return to this point again in \S \ref{sec:err2}.

The red curves, specifically, illustrate the $M_{\rm vir} - L$  relation that is set by forcing the cumulative abundance of dark matter halos more massive than $M_{\rm vir}$ to match the observed cumulative abundance of all galaxies brighter than $L$ \citep[described, for example, in][]{krav04hod,CW08,busha09reion,moster09abund}.  We use the SDSS luminosity function of \citet{blanton05dwarfs} and the halo mass function of \citet[][for WMAP7 cosmological parameters]{tinker08}. To convert from the SDSS bands used in \citet{blanton05dwarfs}  to the $V$-band used in this work, we use the transformation $V=g - 0.59*(g-r) -0.01$ from \citet{jest05sdsstrans}, implicitly assuming all galaxies have average colors.  The line becomes dashed where we have
 extrapolated beyond the luminosity function completeness limit and becomes dotted at large luminosities where
 statistical uncertainties affect our ability to quantify the luminosity function.

  It is encouraging in Figure \ref{fig:abundmatch} that our derived profile matching relation for dMRL-2 (blue, with circles) reveals a similar U-shape as does the abundance matching relation (red).  In particular, our profile matched curve
  reveals a minimum of $M_{\rm vir}/L \simeq 80$ at $M_{\rm vir} \simeq 2 \times 10^{12} M_\odot$ and
$L \simeq 2 \times 10^{10} L_\odot$, reflecting scales where galaxy formation efficiency is maximized.    Similarly, the
abundance-matched curve minimizes at $M_{\rm vir}/L \simeq 80$  
at $M_{\rm vir} \simeq 3 \times 10^{11} M_\odot$ and $L \simeq 4 \times 10^{9} L_\odot$.  This factor of $\sim 6$ agreement is reasonably encouraging,
considering that the minimization of the abundance-matched curve occurs well within the regime where abundance
matching is most affected by baryonic uncertainties.  
Compare the minima to the mass-to-light ratio that would result in the
limiting case where 100\% of each halo's baryons is converted to stars:  $(M_{\rm vir}/L)_{\rm min} = \Upsilon_*/f_{\rm baryon} \simeq 12$ with $\Upsilon_* \approx 2$ set by the
average stellar mass of the E sample in this work ($\Upsilon_*^{\rm E} = 1.89$). The range $1 < \Upsilon_* < 3$ is shown in Figure \ref{fig:abundmatch} as the gray shaded region clearly below any of the matching curves.  The implication is that even for galaxies that are maximally efficient in converting their baryons into stars, some $\sim 70\%$ of their baryons remain unconverted.  
Of course, the inefficiency of baryon conversion into stars is a well-known result of CDM-based comparisons to galaxy luminosity functions.  Nevertheless, it is encouraging that our profile matching analysis seems to imply the same level of inefficiency (on average) without appealing to abundance information in any way.

While the broad-brush agreement between abundance matching and profile matching is encouraging, clearly distinct differences are present for dMRL-2.  There could be several explanations for this.  The most straightforward is that our profile matching $M_{\rm vir}/L$ relations are applicable to dispersion-supported galaxies, while abundance matching applies to galaxies of all types.    This is particularly important in the mass range $M_{\rm vir} \simeq 10^{10-13} M_\odot$ where the population of disky late-type galaxies become much more important relative to spheroidal early-types as mass decreases.  The star forming galaxies will have higher luminosities (lower $M_{\rm vir}/L$) than their pressure-supported/passive counterparts at the same $M_{\rm vir}$, and it is only this latter category that is reflected in our profile matching data set.  Hence, if the star formation efficiency peaks at a different mass for early-type galaxies than late-types, the two methods will give different results for the galaxies in this mass range.   

Additionally, at the bright end, abundance matching typically matches the largest dark matter halos to bright E galaxies.  Thus they do not include the more diffuse, harder to measure intra-cluster stars.  We have included the full CSph light, and therefore the profile matched relation has a larger $L$ at fixed $M_{\rm vir}$ (or lower $M_{\rm vir}/L$).   

With this in mind, it is important to note that at the cluster scale, direct object-by-object comparisons of the measured efficiency \citep{gonz07barycen} is complimentary to the scaling relation approach for comparison to galaxy formation models.  Further, it is  possible to directly compare lensing-based mass estimates to the stellar mass \citep[e.g.][]{zar08eqgal}.  With a large enough sample, this could potentially determine whether there is a discrepancy in either abundance matching or profile matching, although the abundance matching estimates are rather uncertain at these mass ranges due to the impact of small numbers of large clusters (discussed above).  However, because clusters are, by nature, systems where the subhalos/lower-luminosity galaxies are near the peak of efficiency, the host halo of a cluster will always be significantly above the peak.  Thus, this scale cannot probe the mismatch at peak efficiency.  As larger lensing samples at lower masses become available, however, it may be possible to perform direct comparisons at those scales.

The disagreement between abundance matching and profile matching could be further influenced by the use of a luminosity function instead of the $M_*$ mass function.  Because the luminosity function varies depending both on galaxy type (and thus, color) and choice of band, it could bias the inferred abundance matching scales differentially for different galaxy types.  This explanation for the difference in Figure \ref{fig:abundmatch} is supported by results such as \citet{moster09abund} that find a characteristic scale in the $M_{\rm vir} - M_*$ relation at $M_{\rm vir} \simeq 10^{12} M_\odot$, just where our profile matching efficiency is highest.

Other issues affect our interpretation of the dSph galaxies in our sample.
First, almost all of them are
located within the virial radius of the Milky Way, meaning that their dark matter halos are subhalos,
 which may follow different scaling relations.  We consider the effect of this on our derived relations in the next section.  Also, for the very faintest galaxies, we are approaching a regime where surface brightness effects could lead to an observational bias to detect only the highest $M_{\rm vir}/L$ galaxies \citep{bull09stealth}.
 
Despite these caveats, Figure \ref{fig:abundmatch} does clearly show similar patterns to those noted in \S \ref{sec:curve}.  On the faint end, $\Upsilon_{\rm vir}=M_{\rm vir}/L$ shows a much steeper dependence on dark matter mass (this time $M_{\rm vir}$), while the CSph on the bright end are much more sensitive to $L$.  This continues to suggest the dark matter halos are of greater importance for dSphs, while Es and CSph scalings are more controlled by baryonic physics.

A final intriguing property of the profile matching scheme is that there is a built-in consistency check for monotonicity 
in the $M_{\rm vir} - L$ relation.  Specifically, if the $M_{1/2}$ vs. $r_{1/2}$ relation is anywhere 
shallower than the $M_{\rm halo}(r)$ profile it is matching, then the assumption of a monotonic, one-to-one mapping from averaged halo mass (and density profile) to averaged galaxy luminosity must break down.  The fact that the model used here does \emph{not} have this problem implies self-consistency, although it does not guarantee this property in the actual universe.  Clearly, given the size of the measurement errors (see below) the data at this point are not
accurate enough to determine whether or not the relation becomes shallow enough to make the mapping double valued
over a small $r$ range.  We note, however, that if we only consider the smallest (magenta, dSph) galaxies 
($r_{1/2} \lesssim 1$ kpc), the  relation appears consistent with $M_{1/2} \propto r_{1/2}^2$.  For $r<<r_s$ (true for most of the dSphs here), NFW halos obey $M_{\rm halo} \propto r_{1/2}^2$, so the profile matching is just at the limit of monotonicity in the relevant halo mass range \citep[see,][for related discussions]{walker09,wolf09}.  We return to this issue in the next section.

\subsection{Uncertainty and Scatter in the $M_{\rm vir}-L$ relation.}
\label{sec:err2}

Profile matching to the fundamental curve provides a potentially strong constraint on galaxy formation models,
and in principle this method provides a means to test whether or not there is an average, monotonic
$L - M_{\rm vir}$ relation between galaxy luminosity an halo mass, and to investigate the degree of scatter about this relation. Unfortunately, this level of precision testing is hindered by several uncertainties.  First, as discussed in 
\S \ref{sec:err}, there is observational
uncertainty that affects our ability to measure the scatter about and underlying shape of the fundamental curve.  
Second, there is theoretical uncertainty in the {\em average} mapping between an inner mass  
$M_{1/2}^{\rm DM} = M_{\rm halo}(r_{1/2})$  and halo virial mass, which is particularly difficult (and somewhat ill defined)
for the dSph population we consider because they are subhalos.  Finally, even in the limit where the theoretical mapping
between the {\em average} $M(r)$ profile and $M_{\rm vir}$ is perfect, there is a well-known scatter in halo
profiles at fixed mass \citep{Jing00,bul01,Wech02,BK09} and this
 imposes a limiting cosmic scatter in the map between $M_{1/2}$ and $M_{\rm vir}$.
We discuss how all of these issues affect the $M_{\rm vir} - L$ relation in what follows.

\begin{figure*}[bth!]
    \includegraphics[width=0.33\textwidth]{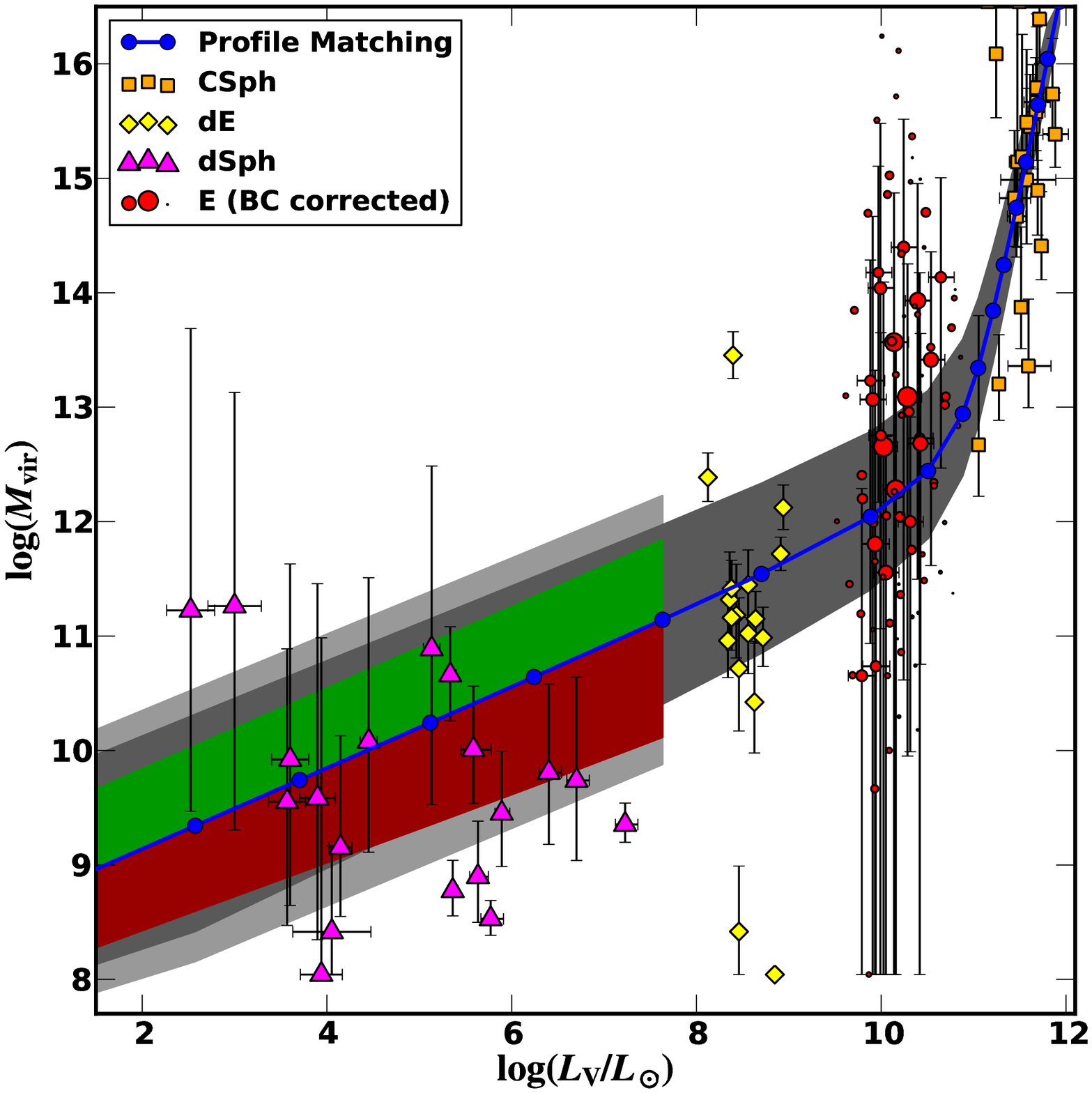}
      \includegraphics[width=0.33\textwidth]{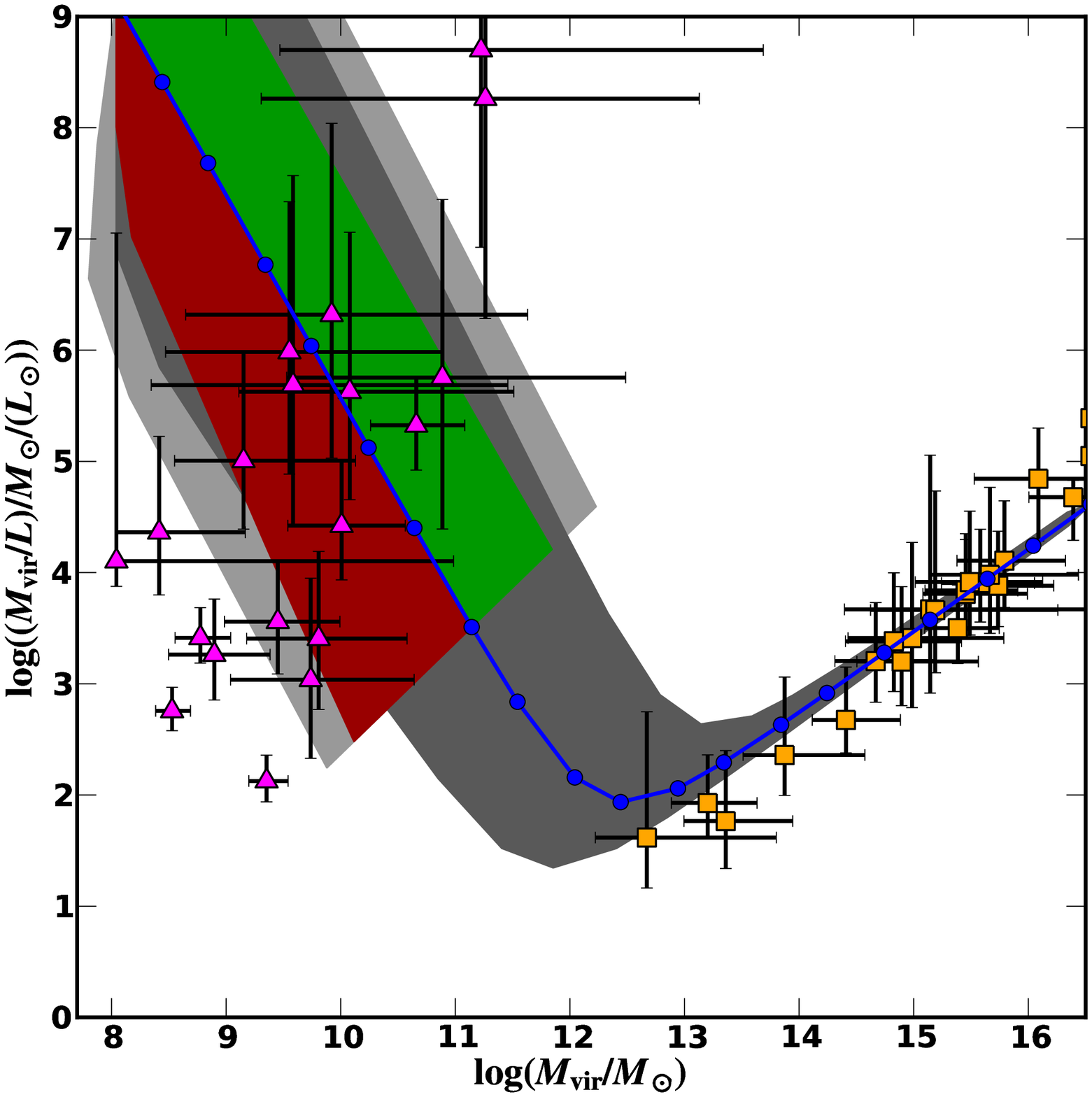}
          \includegraphics[width=0.33\textwidth]{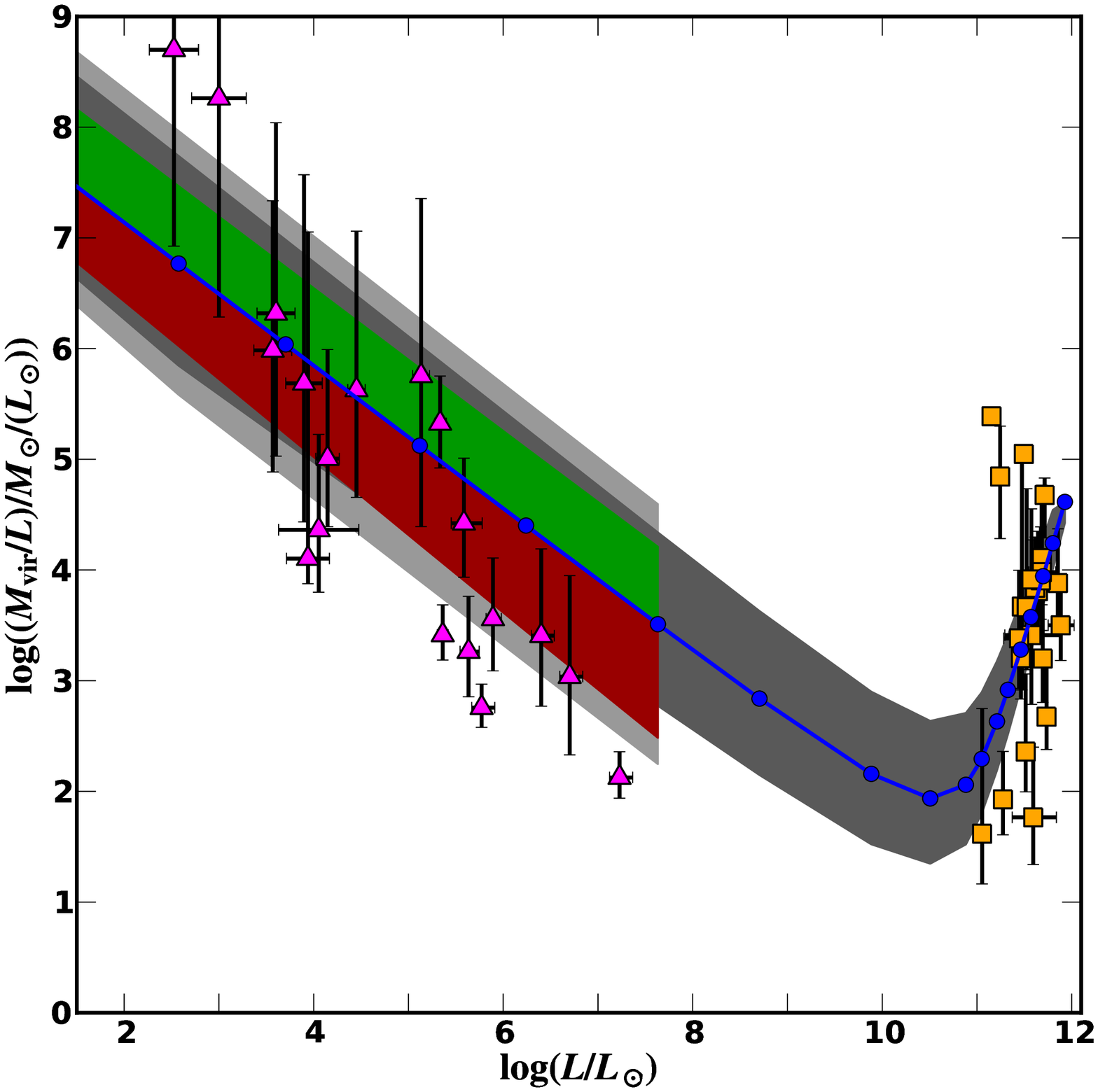}
\caption{
Point-to-point scatter about the median $M_{\rm vir} - L$ relations:
V-band luminosity against virial mass (left),
virial mass-to-light ratio against virial mass (middle), and virial mass-to-light ratio against luminosity (right).
The blue line with points is the median relation for the dMRL-2 model.  The width of the shaded band
is the  minimum galaxy-to-galaxy scatter that we expect in the limit where virial mass correlates with
$L$ with zero scatter.  The band is {\em not} the error in the $M_{\rm vir} - L$ relation, but rather the
minimum derived scatter set by a fundamental cosmological uncertainty in the mapping between $M(r_{1/2}) = M_{1/2}^{\rm DM}$ and $M_{\rm vir}$. On the faint end, where the galaxies in our sample are in sub-halos, the inferred mass can be scattered upwards due to uncertainty in infall redshift (green inner shaded region), and downward due to mass loss (red inner shaded region).  Adding these effects in quadrature with the cosmic variance (middle dark gray shaded region) results in the full cosmological scatter (outer light gray shaded region). 
Points represent the objects described in \S \ref{sec:dat}, with virial masses inferred by inverting the grid of Figure \ref{fig:massplot} and error bars  due to the effect of the {\em observational} error bars on the mapping, including a factor of 2 uncertainty in $M_*$ for dEs and Es. For clarity, error bars are only shown for the 20 E galaxy bins with the largest number of galaxies.  The $M_{1/2}^{\rm DM}$ assumed for the Es has been corrected for baryonic contraction using the estimate described in \S \ref{sec:halomatch}.  For the CSphs, it is important to note that the vertical error bars are likely underestimated due to the uncertain effect of using the cluster galaxies instead of the intra-halo light in measuring the velocity dispersion.  The data point color and point-type scheme matches that of the previous figures (see \S \ref{sec:dat}).   Intrinsic scatter in the $M_{\rm vir} - L$ relation can only be detected in cases where the observational error bars are small enough to detect a dispersion smaller than the shaded band. We see that the population of dSph galaxies is close to this level now.}
\label{fig:mlscatter}
\end{figure*}

\begin{figure*}[bth!]
    \includegraphics[width=0.33\textwidth]{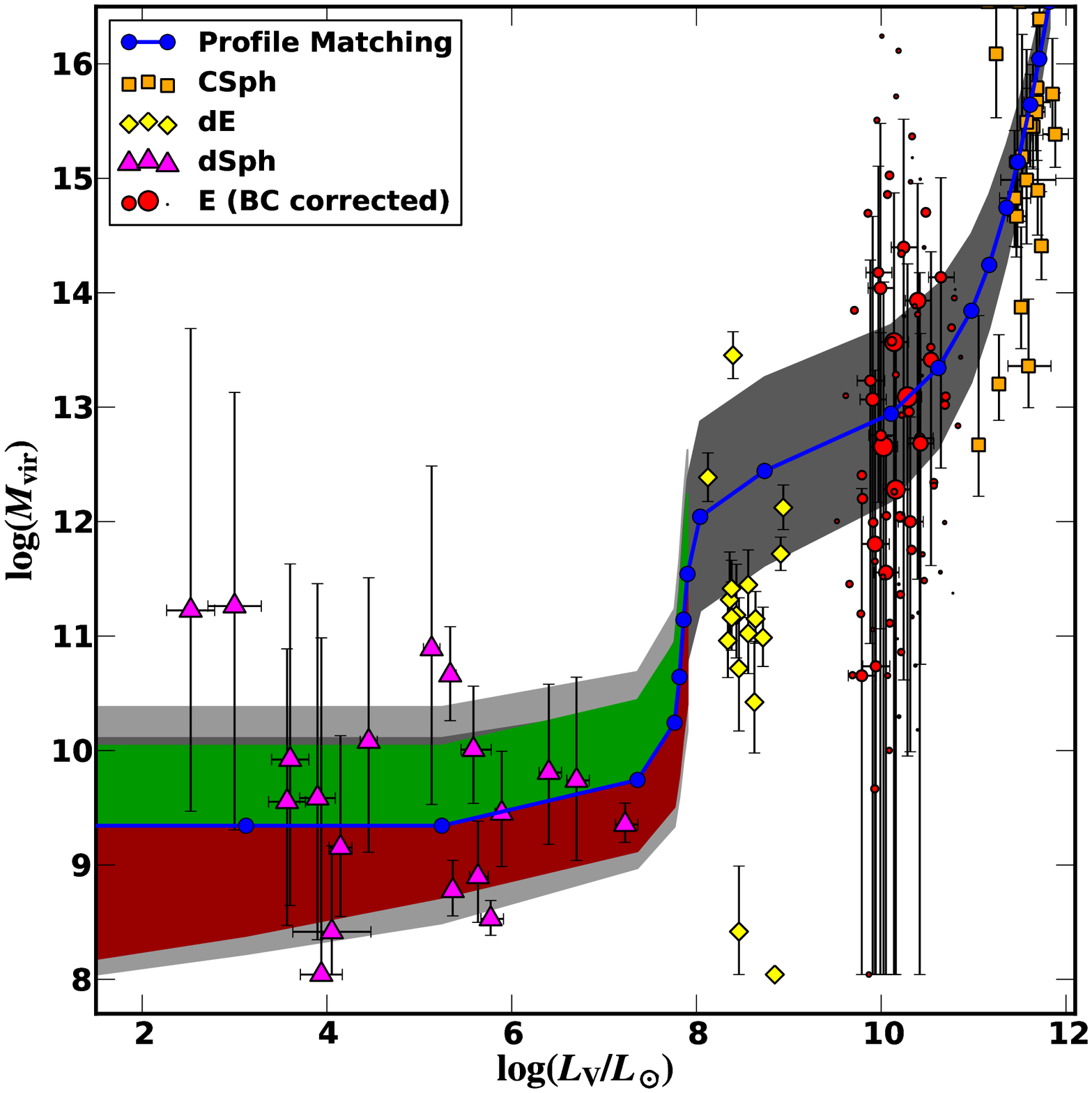}
      \includegraphics[width=0.33\textwidth]{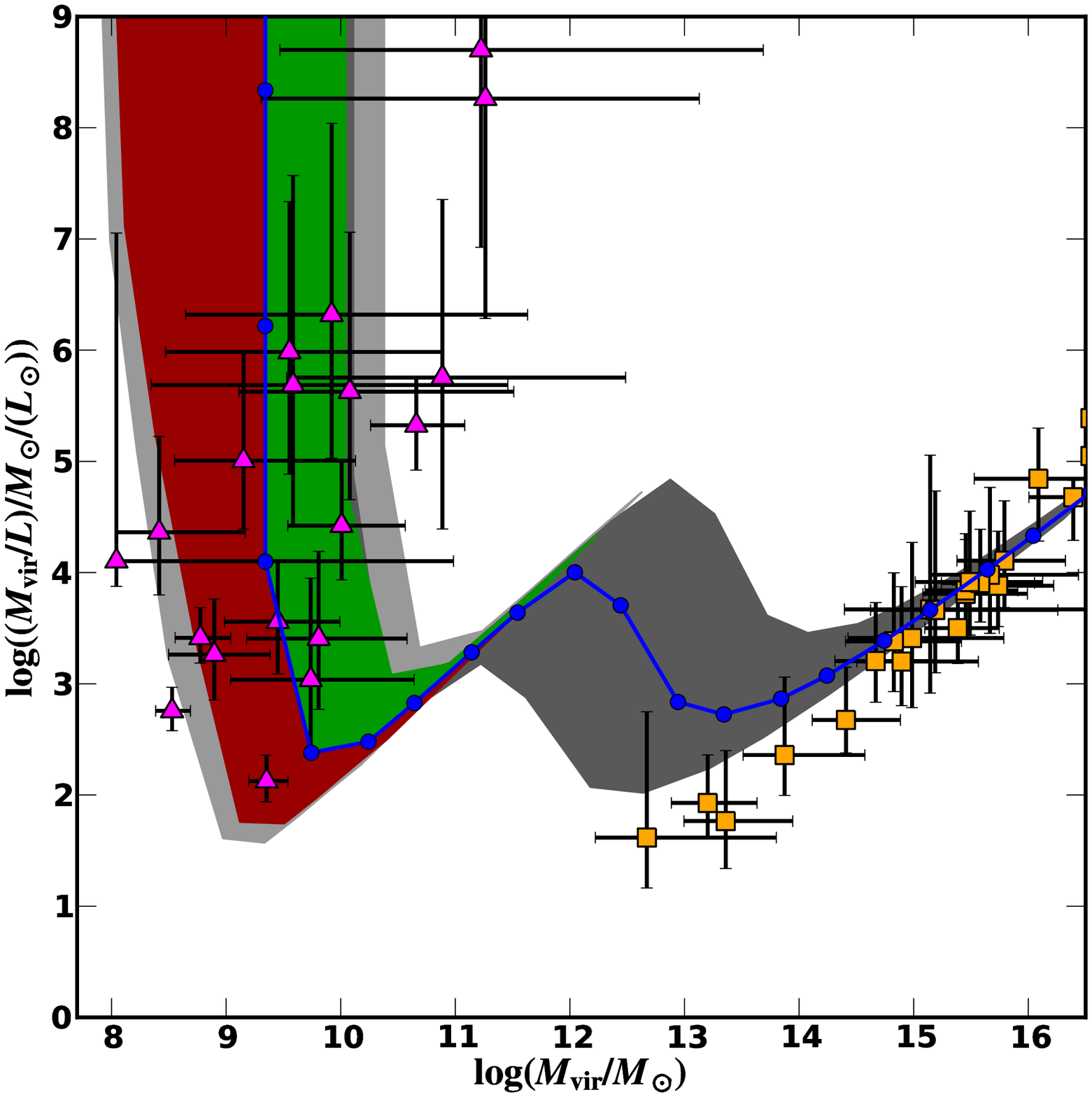}
          \includegraphics[width=0.33\textwidth]{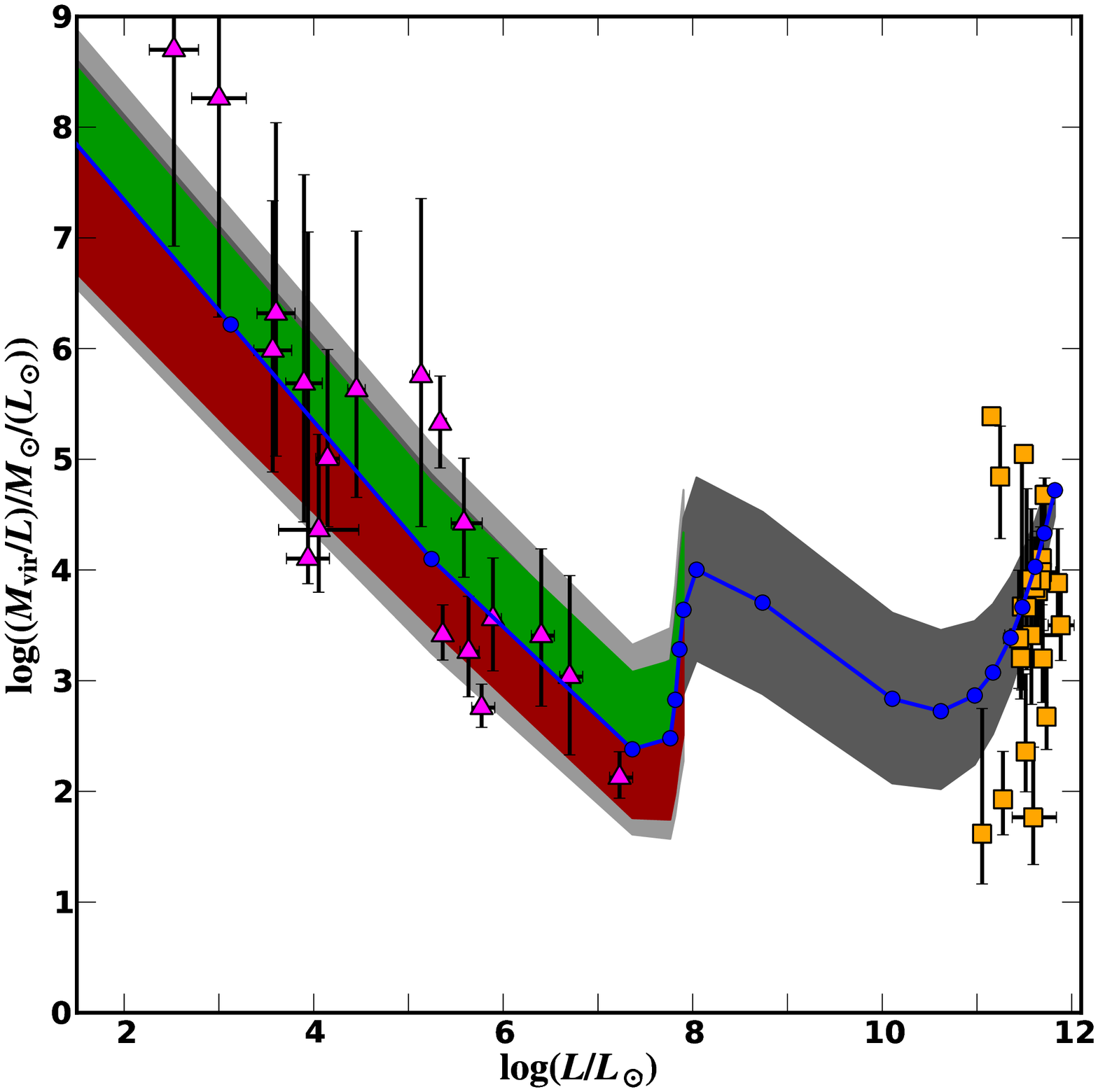}
\caption{Same as Figure \ref{fig:mlscatter}, but using the dMRL-3 model to compute the profile matched relation (blue solid line). The nature of the ``bump'' feature is discussed in \S \ref{sec:halomatch}.}
\label{fig:mlscatter_altmodel}
\end{figure*}

Figures \ref{fig:mlscatter} and \ref{fig:mlscatter_altmodel} provide visual presentations of the observational and theoretical uncertainties in the profile matching relations for $M_{\rm vir}$ vs. $L$ (left) and the equivalent implied relations of $M_{\rm vir}/L$ vs $M_{\rm vir}$ (middle) and 
$M_{\rm vir}/L$ vs. $L$ (right).   Starting with observational uncertainties, the error bar on $M_{\rm vir}$ for each data point is estimated by offsetting the observables by their $1\sigma$ errors in $M_{1/2}^{\rm DM}$ and $r_{1/2}$, and performing the profile matching for each data point individually (using the mean fundamental curve relation for Model 3).  For the Es, we use the error bars adopted in the previous section (factor of 2 on $M_*$).  We note that this this implies very large errors on $M_{1/2}^{\rm DM}$ for the E (and dE) galaxies, because these are the systems for which $M_*/2$ is closest to $M_{1/2}$, and hence the possible error in $M_*$ has the largest effect on $M_{1/2}^{\rm DM}$.  This large uncertainty in $M_{1/2}^{\rm DM}$ maps to an even larger (relative) uncertainty in $M_{\rm vir}$.  Figures \ref{fig:mlscatter} and \ref{fig:mlscatter_altmodel} are distinguished by use of the dMRL-2 and dMRL-3 models, respectively.

Next we consider the cosmological scatter in the dark matter mass enclosed within a given radius for an ensemble of halos with identical virial masses.      For field halos, this
scatter can be accounted for by the scatter in the concentration-mass relation for dark halos, which is approximately
log-normal in concentration with a variance of $\Delta \log(c) = 0.14$ \citep{Wech02}.
In principle, this cosmic scatter provides a lower limit on point-to-point scatter that can be measured in 
a profile matched $M_{\rm vir}-L$ relation.  We illustrate the magnitude of this cosmic scatter by
the middle (dark gray) shaded band, which traces our best-fit relation (shown as a solid blue line connecting blue circles) in
each panel. We see that this cosmic variance is particularly important for the smallest galaxies.
This cosmic variance scatter is the minimal possible scatter expected for galaxies in $\Lambda$CDM.   Even if
galaxy properties tracked virial mass in a precisely one-to-one fashion, they would scatter about the profile matching 
relation with at least this amplitude.   \footnote{In principle, if galaxy luminosity had a secondary dependence on halo concentration, then the covariance could act to reduce the cosmic scatter from profile matching, but this seems tuned and unlikely.}

An additional component of scatter and uncertainty must be considered for the dSph galaxies -- because they are satellites of the MW, their dark matter halos are subhalos, and hence do not obey the same scaling relations as field halos
\citep[e.g][]{bul01,springel08aquarius}.  More specifically, it is inappropriate to speak of a virial mass for a subhalo, because subhalos
tend to be tidally truncated at radii that are smaller than the virial radius they had when they were first accreted. 
A more meaningful mass to be associated with each dSph is its halo's virial mass at the time it was accreted.
It is this mass, the virial mass at accretion, that would most likely show a strong correlation with galaxy luminosity.

Two competing effects may act to modify standard (field) mapping between inner mass and virial mass.
First, at fixed virial mass, a halo at higher redshift will tend to be denser at a fixed physical radius than a halo
of the same virial mass at a later redshift (because the virial density scales roughly with the density of the universe).
Therefore, if a subhalo was accreted at some high redshift (e.g. $z=3$) and it experienced no mass loss in its central
regions (unlikely) then our virial mass estimates are biased high.  
The lower (red) shaded region in the $L < 10^7$ L$_\odot$ band of Figure \ref{fig:mlscatter} illustrates
the degree by which the median relation would need to be shifted down in order to account for
a $z \le 3$ accretion that experienced no mass loss within its central region after accretion.  The lower edge of the
red band corresponds to the relation expected if \emph{all} dSphs were accreted at $z=3$ with no mass loss.

The second, competing processes that adds uncertainty to profile matching estimates for subhalos is tidal mass loss.
Halos tend to lose mass at all radii after they are accreted, and this acts to decrease their central density
for a fixed virial mass at accretion.  The cosmological simulation of
 \citet{BK09} shows that the median subhalo at $z=0$ in a Milky-Way-type host has
 lost $75 \%$ of its initial {\em total} mass,  while  $\sim85\%$ of subhalos in  have lost 
 $< 90 \%$  of  their initial \emph{total} mass  (Boylan-Kolchin 2010, private communication).   
However, the mass loss is far less significant in the inner regions we are probing here \citep{K04,pen08,wetzel10,pen10}.
The simulations of \citet{BJ05} show that a $75\%$ ($90\%$) loss of {\em total} mass, results in a mass loss
fraction within the inner 300 pc of only $20 \%$ ($40 \%$) -- where $r_{1/2} = 300$ pc is the median half-light radius for our dSph sample.   For the mass range of relevance $M_{\rm vir} \propto M_{300}^{3.3}$ \citep{bull09stealth}, which implies that our fiducial $M_{\rm vir}$ determination from field halo profile matching would be under-estimated
by a factor of $(0.8)^{-3.3} \sim 2$ for median subhalo mass loss, and by a factor of
 $(0.6)^{-3.3} \sim 5$ in the case of 90\% total mass loss.  Thus, in Figure \ref{fig:mlscatter} we include an upper (green) shaded region corresponding to a factor of 5 increase in the inferred $M_{\rm vir}$, as a conservative estimate of the maximal scatter. This treatment is conservative because we expect that systems with the most mass loss will also have been
 accreted earlier, and therefore to have had higher virial densities overall.  This offsetting effect has been ignored in
 the upper green shaded band.
 
Of course, if we knew
the redshift of accretion and orbital trajectory (including mass loss)
of each dSph in our sample, we could perform the profile matching in a 
more exacting way, but this is not practical with present data due to the uncertainties in the orbits of the MW satellites \citep{lake10dwarforbits}.  Therefore, we have added both the effects in quadrature to the cosmic variance error band in Figure \ref{fig:mlscatter} in order to derive a limiting scatter estimate shown as the outer light gray regions in Figure \ref{fig:mlscatter}.

Thus, the shaded bands about the average relations in Figure \ref{fig:mlscatter} can be thought of as a limiting theoretical scatter about the relation.  In principle, if the data at a particular scale scatter about the relation with a larger variance than indicated by the shaded band, then this would be indicative of intrinsic scatter in the $M_{\rm vir} - L$ relationship.  This then implies that the secondary scalings in galaxy formation (e.g. 2-D scalings such as the fundamental plane) can be fit to the data set to provide useful information.  Conversely, if the scatter (including observational errors) is consistent with the theoretical scatter, the secondary scaling relations cannot be measured at that scale.

The possibility of detectable intrinsic scatter is is particularly interesting at the faint end, where it has been noted that despite the wide ranges of luminosities, the MW dSphs appear to have similar halo masses, albeit with large scatter \citep{stri08nat,wolf09} -- this could be due to scatter (observational or intrinsic) masking a weak relation, scatter in halo mass about a new scale in galaxy formation, selection effects (e.g. the stealth galaxies' influence as discussed in \S \ref{sec:halomatch}), or some as yet unknown effect.   This scale appears particularly strongly in \ref{fig:mlscatter_altmodel} due to the preferential fitting on the dSph. These data also admit a steepening power law instead of a true scale at the low-mass regime as suggested by \citet{krav09drev} to match the dSph luminosity function, so we plot this relation in Figure \ref{fig:abundmatch}.
 
We note in Figure \ref{fig:mlscatter} that there is a systematic offset for the bright dSphs.  This is primarily due to the tension between fitting the RM relation for the dE and the bright dSphs with a single power law, as is used for dMRL-2.  In Figure \ref{fig:mlscatter_altmodel}, this offset is essentially gone, as the fit in the RM relation is tailored to fit best for the dSphs.  This comes at the price of a poorer fit for the other galaxies, however, as well as an anomalously low $M_{\rm vir}/L$  apparent in Figure \ref{fig:abundmatch} (green-dashed line in lower-right panel).  It is unclear if this tension is due to problems with $\Lambda$CDM accounting for the existence of galaxies in the halos of the bright dSph, evolutionary effects on subhalos (as discussed above), or the influence of baryonic contamination of $M_{1/2}$, which is unaccounted for in our analysis of the dSphs.

 Unfortunately, as is clear from comparing the data points to the shaded band in Figure \ref{fig:mlscatter}, the observational uncertainty is still slightly too large on the faint end to determine if there is significant \emph{intrinsic} scatter about the fundamental curve, although the data are close. Similarly, while the very faintest galaxies show deviation from fundamental curve in a way consistent with a new scale of flat $M_{\rm vir}/L$, this level of deviation is not statistically significant.  Similar uncertainties likely apply to M31 satellites, making it difficult to interpret the possible existence of an offset \citep{kal10}.   Marginal improvements in data quality may be enough to shed light on these questions, however, as observational errors could be brought to the level of cosmological scatter.  Furthermore, the predicted existence of far more faint dSphs in the Local Group to be detected in upcoming surveys \citep{toll08,martin09cubs,bull09stealth} provides hope that this degeneracy between intrinsic and observational scatter may be  broken by sheer statistics.  Nevertheless, the current data are not good enough to definitively address this question.

There is also hope on the bright end.  Interestingly, the most massive, luminous objects are the ones that face the least
cosmological scatter associated with the profile matching technique.  As can be seen in Figure \ref{fig:massplot}, as one travels
along the fundamental curve projection to large values of $r_{1/2}$ and $M_{1/2}^{\rm DM}$, 
the associated $M_{\rm vir}$ determinations become more cleanly defined.  Unfortunately, it is in this regime where
our inability to determine CSph velocity dispersions limit the ability to cleanly determine $M_{1/2}$.

\section{Conclusions}
\label{sec:conc}

We have examined the scaling relations for a broad collection of spheroidal stellar systems in an intrinsic 
MRL space of half-light mass ($\Mhalf$; Equation \ref{eqn:Mh}), half-light radius ($\rhalf$), and half-luminosity ($L_{1/2}$).  
These MRL coordinates
are a theoretically-motivated  transformation of
the familiar fundamental plane variables and can serve as a 
 bridge between direct observables and the predictions of  galaxy formation models.  
The latter is facilitated by considering an alternative space we refer to as dMRL space.
In dMRL space, the mass variable is $\Mhalf^{\rm DM}$ -- the dark matter mass within $\rhalf$
rather than the dynamical mass.

Our main findings are as follows.

\begin{enumerate}  

\item All spheroidal {\em galaxies}---stellar systems with their own dark matter halos---track a 1-D fundamental curve through MRL space.  This curve is visualized in 3-D in Figure \ref{fig:fcurve3d} and represented analytically
in Equations \ref{eqn:arctanrvsl} and \ref{eqn:arctanrvsm} (with parameters from Table \ref{tab:curveparams}).     
The fundamental mass-radius-luminosity relation transitions from $M_{1/2} \propto r_{1/2}^{1.44} \propto  L_{1/2}^{0.30}$ for the faintest dwarf spheroidal (dSph) galaxies to $M_{1/2} \propto r_{1/2}^{1.42} \propto  L_{1/2}^{3.2}$ 
for the most luminous cluster spheroids (CSphs). This $\rhalf-\Lhalf$ scaling (MRL-2 model) is a good fit for the dSphs if we take into account the fact that the lowest luminosity dwarf galaxies suffer from surface brightness incompleteness (which biases the sample towards smaller $\rhalf$).  If we ignore this bias, the raw empirical relation (MRL-1 model) gives $M_{1/2} \propto r_{1/2}^{1.44} \propto  L_{1/2}^{0.86}$ on the faint end.

\item  Dwarf ellipticals (dEs) and normal ellipticals (Es) inhabit the transition regime between the limiting power laws, where the dynamical mass-to-light ratio within $\rhalf$ is minimized at $\Upsilon_{1/2} \simeq 3$.  The dynamical mass as a function of $\rhalf$  transitions quite abruptly  as the galaxies become baryon-dominated (see Figure \ref{fig:fcurve2d}).  When we subtract out the baryonic component with estimates for the stellar mass (although these are subject to uncertain systematic errors), the relation is better fit by a power law, particularly when we include an estimate for the effect of baryonic contraction (see the inset of Figure \ref{fig:massplot}). The inferred slope for the $\rhalf - \Mhalf^{\rm DM}$ relation is $\Mhalf^{\rm DM} \propto r_{1/2}^{2.32}$, slightly steeper than the  $M \propto r^2$ relation that has been discussed in the literature \citep{gentile09,napo10cendm,walker10dsphsp}.

\item Globular clusters (GCs) and ultra-compact dwarfs (UCDs) do not follow the fundamental curve relation.
Instead, GCs and UCDs  inhabit overlapping/connecting 
regions in MRL space that resemble sections of mass-follows-light planes near  $M_{1/2} = 3 \,  L_{1/2}$,
as illustrated in Figure \ref{fig:sepplot}. See Equation \ref{eqn:mlrsepeq} for the exact form of the plane that separates this GC locus from the dSph portion of the fundamental curve.
 Note that the UCDs in our sample exhibit a small ``tilt" away from the mass-follows-light plane,
while GCs exhibit no such tilt -- thus it cannot be ruled out that UCDs are a part of the galaxy sequence, but are intrinsically rare
in the region where they meet the fundamental curve. 
However, dSphs separate distinctly from GCs and UCDs in MRL space, implying that if UCDs are actually embedded in dark matter halos, an irreducible
dichotomy exists in galaxy formation.

\item The fundamental curve relation in dMRL space allows us to connect galaxies to their dark matter halos
via an approach we call profile matching.  Specifically, at each luminosity, an {\em average} galaxy sits in a 
specific point in the $\Mhalf^{\rm DM} - \rhalf$ plane.  This mass-density 
point can be mapped to an {\em average} dark matter halo virial mass, as illustrated in Figure \ref{fig:massplot}.  
While we assume standard NFW halos in $\Lambda$CDM, this technique is easily adaptable to any dark matter halo type that can be cast as a single-parameter family.
In the end, we can construct relationships between luminous galaxy properties and their dark matter halo masses.
This profile matching technique for deriving the $M_{\rm vir} - L$ is most accurate at the high and low luminosity extremes (where dark matter fractions are highest) and is therefore quite complementary to statistical approaches that rely on having a well-sampled luminosity function.

\item Independent of any global abundance or clustering information, we find that (spheroidal) galaxy formation needs to be most efficient in $\Lambda$CDM halos of virial mass $M_{\rm vir} \simeq 10^{12} \, M_\odot$ and to become {\em sharply}
inefficient in masses smaller than $M_{\rm vir} \lesssim 10^{10} \, M_\odot$.  
On the other hand, the inefficiency of galaxy formation seems occur more gradually as halos become more
massive than $M_{\rm vir} \simeq 10^{13} \, M_\odot$.  Rather, the inefficiency sets in  sharply in luminosity
at $L \simeq 10^{11} \, L_\odot$.  These results are qualitatively consistent with the expectations of abundance matching (see Figure \ref{fig:abundmatch}), although only if we use models that account for surface brightness selection effects on the faint end (dMRL-2 and dMRL-3).   
The sharpness of the transition with $M_{\rm vir}$ on the faint end may imply the dark matter halo or potential depth drives
scaling relations for low-mass galaxies, while the stronger dependence on $L$ on the bright end suggests baryonic physics controls the massive galaxy regime.

\item Object-by-object scatter about the $M_{\rm vir} - L$ relation remains very difficult to quantify. Nevertheless, 
despite the large
theoretical uncertainties associated with our profile matching technique at the low-mass end, the observational data
for dSphs 
are almost to the point where we can explore intrinsic scatter about this relation in the smallest systems.  On the other
hand, the theoretical uncertainty in the mapping between points in the $\Mhalf^{\rm DM} - \rhalf$ plane and halo virial
is much smaller on the scale of CSphs, so there is promise at the bright end from that respect.  Unfortunately, 
stellar velocity dispersion for CSphs remain very difficult to obtain directly.  A better approach would be to consider
alternative mass-radius measurements for CSphs (based, for example, on X-ray studies) as has recently been explored
by \citet{tgkly10}.

\end{enumerate}

We close by mentioning that the existence of a fundamental curve in MRL space is not out of line with an understanding 
that galaxy properties show strong correlation with a single parameter
 \citep[see, e.g.][for similar results on an HI-selected sample]{disney08simplegals}.  Nevertheless, this fact does
\emph{not} imply that all galaxies belonging to a given evolutionary sequence are completely or even primarily 
controlled by a single parameter -- only that their first-order scaling relation 
is characterized by a single parameter when the galaxy properties are averaged. 
Our viewpoint is rather that the MRL relation presented above provides a 
useful bridge between observational properties and theoretical models.  At the very least, models should be able to
reproduce the 1-D scaling relation presented.  Some guidance to that aim is provided by our dMRL-inspired 
profile matching, which seeks to unite galaxies across a space of virial mass, stellar luminosity, and stellar radius
self-consistently.

We wish to acknowledge Sandra Faber, Andrey Kravtsov, and Chris Purcell for helpful discussions.  We also thank the anonymous referee for a useful and clarifying report.  This work was supported by the Center for Cosmology at UC Irvine. E.J.T. was supported by the UCI Physics \& Astronomy GAANN fellowship.

\bibliography{paper}{}
\bibliographystyle{hapj} 

\appendix

\section{Analytic Fit to Halo Matching Relations}
\label{apx:matchfit}

In order to provide an analytic description of the derived $M_{\rm vir}$ relations for our fiducial (dMRL-2) model results, we perform a least-squares fit of $y$ vs. $M_{\rm vir}$ 
 for each of $y = L_{1/2}$, $r_{1/2}$,  $M_{1/2}$, and $(M_{\rm vir}/L)$ using the same fitting form as Equation \ref{eqn:arctanrvsl}
 
\begin{align} 
\log\left(\frac{y}{y_0}\right) =  \Mutil \, \frac{A + B}{2} + \left[ S - \Mutil (A - B) \right] \frac{\arctan(\Mutil / W)}{\pi}.
\label{eqn:matchfit}
\end{align}
Here, $\Mutil \equiv \log (M_{\rm vir}/M^0_{\rm vir})$ defines a characteristic virial mass scale $M^0_{\rm vir}$
at $y = y_0$ and $W$ sets the width of the transition from
$y \propto M_{\rm vir}^A$ and
$y \propto M_{\rm vir}^B$ at small and large $M_{\rm vir}$, respectively, and $S$ sets the offset in $\log(y)$ over the transition region.
Each of the fit parameters $A$, $B$, $W$, $S$, $M^0_{\rm vir}$, and $y_0$ are provided in Table \ref{tab:matchtab} for 
our four $y$ relations, corresponding to the four panels of Figure \ref{fig:matchplots}.

We find that the $\Mhalf^{\rm DM}$ vs.  $M_{\rm vir}$ and $\rhalf$ vs. $M_{\rm vir}$  relations for dMRL-2 are also very well characterized
by power-laws.  Specifically we find
\begin{equation}
M_{1/2}^{\rm DM} \simeq \left(\frac{M_{\rm vir}}{1.35 \times 10^{5}  M_{\odot}}\right)^{1.36} \, M_\odot \, ,
\end{equation}
and
\begin{equation}
r_{1/2} \simeq \left(\frac{M_{\rm vir}}{2.17 \times 10^{11} M_{\odot}}\right)^{0.59} \, {\rm kpc}.  
\end{equation}
The $L$-to-$M_{\rm vir}$ relation, meanwhile, can be approximated on the faint end as $L \propto M_{\rm vir}^{2.84}$ and flattens on the bright end to $L \propto M_{\rm vir}^{0.26}$.

\begin{deluxetable}{|c|cccccc|}
\tablecolumns{7}
\tablecaption{Profile Matching Model Parameters for Equation \ref{eqn:matchfit}}
\tablehead{
\colhead{$y$} &
\colhead{$y_0$} &
\colhead{$M_{\rm vir}^0 / M_\odot$} &
\colhead{$A$} &
\colhead{$B$} &
\colhead{$W$} &
\colhead{$S$}
}

\startdata

$M_{1/2}^{\rm DM}$ & $2.03 \times 10^{12} M_\odot$ & $2.85 \times 10^{14}$ & 1.38 & 1.31 & 2.40 & 0 \\
$L_{1/2}$ & $8.95 \times 10^{9} L_\odot$ & $1.78 \times 10^{12}$ & 2.84 & 0.26 & 0.71 & 0 \\
$r_{1/2}$ & $70 \; {\rm kpc}$ & $2.85 \times 10^{14}$ & 0.60 & 0.56 & 2.40 & 0 \\
$M_{\rm vir}/L$ & $199 (M/L)_\odot$ & $1.78 \times 10^{12}$ & -1.84 & 0.74 & 0.71 & 0

\enddata

\label{tab:matchtab}
\end{deluxetable}

\section{Alternative Data Projections}
\label{apx:altdata}

\begin{figure*}[htb!]
\epsscale{1.15}
\plotone{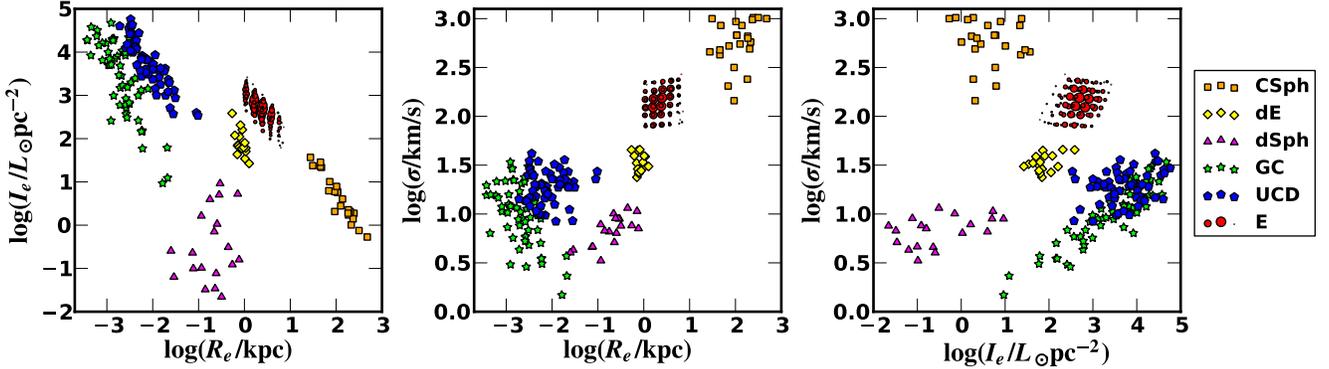}
\caption{Two dimensional projections of the three-dimensional data shown in Figure \ref{fig:fp3d} along the coordinate axes.  The color/symbol code maps to galaxy type as indicated, matching the scheme of Figure \ref{fig:fjplot}.}
\label{fig:fp2d}
\end{figure*}

\begin{figure}[htb!]
\epsscale{0.95}
\plotone{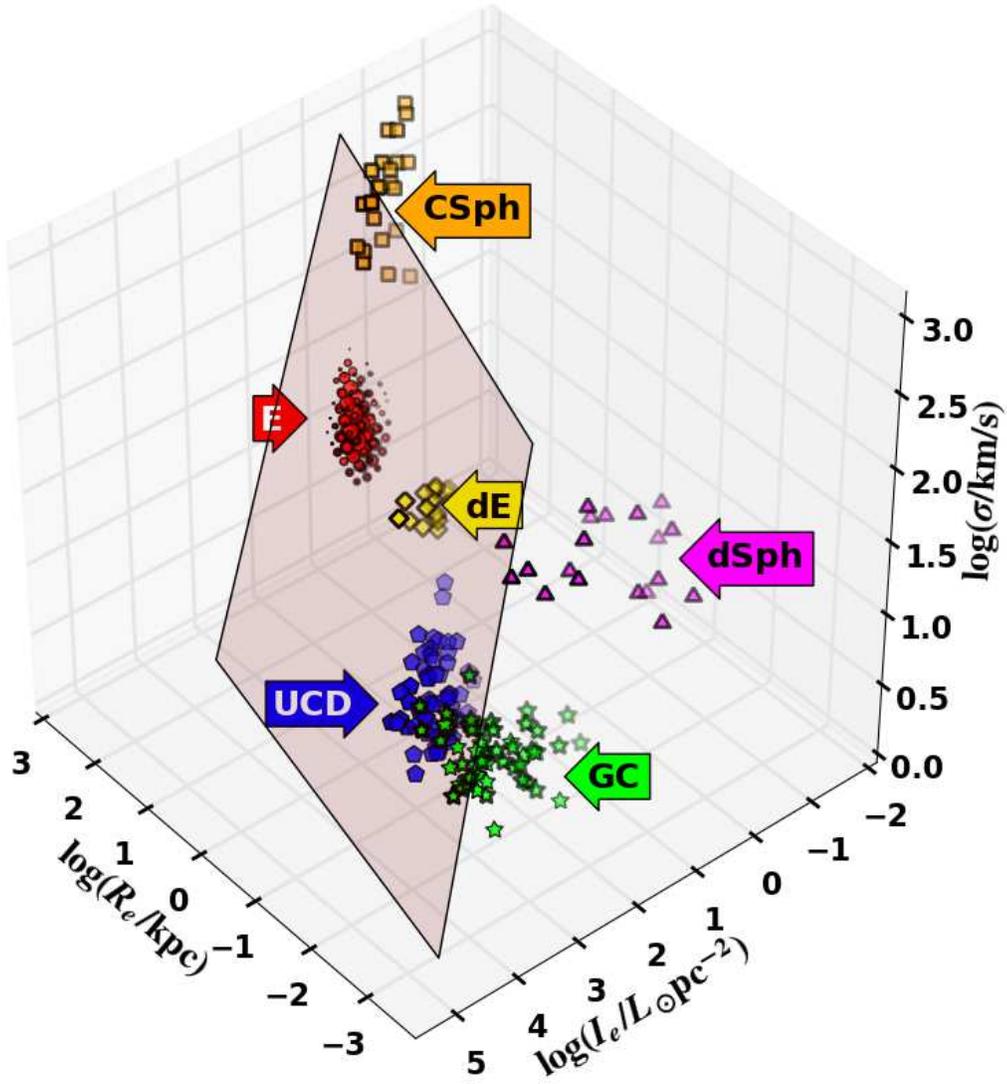}
\caption{ Three dimensional representation of the data set in fundamental plane coordinates of $\log(I_e/L_\odot {\rm pc}^{-2})$, $\log(R_e/{\rm kpc})$, and $\log(\sigma/{\rm km/s})$.      The red transparent plane is the fundamental plane for ellipticals from \citet{graves09ii}, and the color/symbol code matches Figure \ref{fig:fjplot}. A rotating animation of this plot is available at \movieurl.}
\label{fig:fp3d}
\end{figure}

The fundamental plane of bright elliptical galaxies \citep{dj87fp,dressler87dnsig,faber87fp} lies within a 3-D parameter space that consists of the velocity dispersion ($\sigma$), the 2-D half-light (effective) radius ($R_e$), and the surface brightness ($I_e$).  Here we define $I_e$ such that it is the \emph{mean} surface brightness within $R_e$, in units of $L_\odot\,{\rm pc}^{-2}$, although we note that slightly different definitions are sometimes used in the  literature.
The fact that these three variables are direct observables that scale together motivates the
consideration of galaxies in this space.  Two-dimensional projections of our data set (Table \ref{tab:dat}) 
on the fundamental plane axes are shown in the three panels of Figure \ref{fig:fp2d}.  In Figure \ref{fig:fp3d} 
we plot in a 3-D rendering of these same data.  Also shown in transparent red in Figure \ref{fig:fp3d} is the best-fit fundamental plane of \citet{graves09ii}.

From these plots, it is apparent that while the normal elliptical galaxies (E)  lie well within the fundamental plane of \citet{graves09ii}, the CSphs and dSphs lift away from the plane in a non-trivial manner, in contrast to data sets that do not reach those extremes in luminosity \citep[e.g.][]{burbend97}.  However, it has been noted in the literature that the faint end of the fundamental plane (towards dEs) shows curvature up off the plane \citep{zar06fmdwarfs, hyde09curvefp}, and bright-end deviations from the fundamental plane are discussed in further detail in \citet{zar06fman}.  Here we note that the separation from the plane is much more 
significant when the dSph galaxies discovered in the SDSS ($R_e \lesssim 450$ pc) are included alongside the ``classical'' dwarfs, as the SDSS dSphs extend nearly perpendicularly from the fundamental plane. with that in mind, the deviation from the plane is significant far beyond the  scatter in the fundamental plane derived for bright E galaxies.  

The ``tilt'' of the E fundamental plane here can be interpreted in the context of this curvature; the tilt of the fundamental plane is simply the shift of the observational fundamental plane from the expected virial plane (see \S \ref{sec:mlrspace} and the two planes in Figure \ref{fig:mlr3d}).  Curvature off the plane is then just continuation of this tilt past the typical regime of Es.  The tilt in the Es can potentially be driven by a mix of stellar mass-to-light ratio variations and/or variation in the dark matter-to-baryon fraction within the halo of the galaxy in question \citep{cappellari06,bolton07,humph10,napo10cendm,treu10imf,graves10iii}.  It may also be an aperture affect due to dissipation causing a change in the apparent dark matter fraction by packing more baryonic material in the same volume of dark matter halo \citep{robertson06ellscale,hopkins08fpdiss}.  For our purposes, however, it is sufficient to note that the magnitude by which the largest and smallest spheroidal galaxies peel away from the fundamental plane cannot be explained by baryonic effects  -- it can only be addressed in terms of dark matter content, due to the very large mass-to-light-ratios.

\begin{figure*}[htb!]
\epsscale{1.15}
\plotone{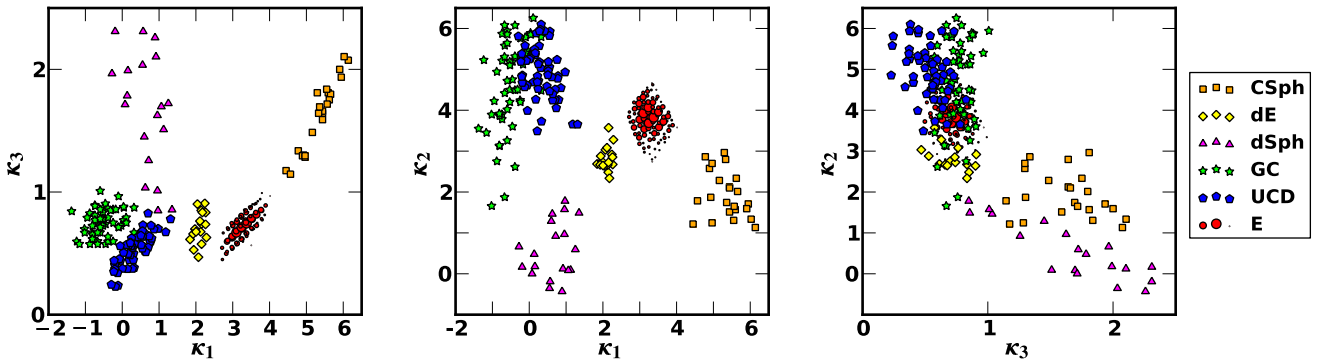}
\caption{Two dimensional projection of the three-dimensional data shown in Figure \ref{fig:fp3d} onto the $\kappa$ space of \citet{b2f1}:  $\kappa_1 \propto \log{M}$, $\kappa_2 \propto \log{\left((M/L)I_e^3\right)}$, and $\kappa_3 \propto \log{(M/L)}$, and $\kappa_1$ and $\kappa_2$ are approximately parallel to the fundamental plane. The color/symbol code maps to galaxy type as indicated, matching the scheme of Figure \ref{fig:fjplot}.}
\label{fig:fpproj}
\end{figure*}

For comparison with other work,  Figure \ref{fig:fpproj} shows the projection of the data onto the kappa ($\kappa$) space of \citet{b2f1}, a coordinate rotation that enables a reasonably physical interpretation 
with $\kappa_1 \propto \log{M}$,  
$\kappa_2 \propto \log{\left((M/L)I_e^3\right)}$, and $\kappa_3 \propto \log(M/L)$ such that $\kappa_1$ and $\kappa_2$ 
 define a plane that is approximately parallel to the fundamental plane for ellipticals.

\begin{figure*}[htb!]
\plotone{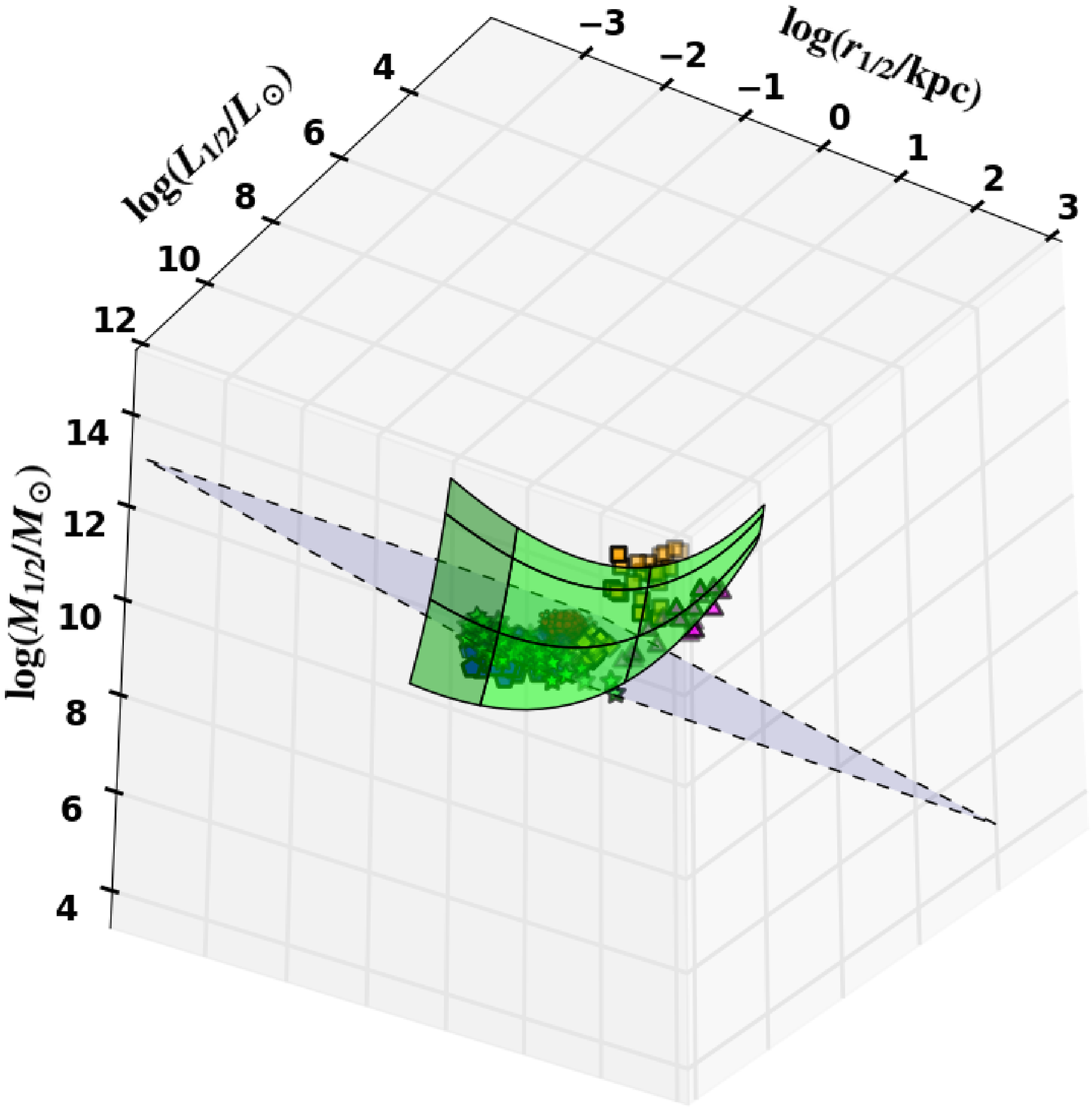}
\caption{Three dimensional representation of the fundamental manifold of \citet{zar06fman} in MRL space (i.e. Figure \ref{fig:mlr3d}).  The transparent (green) manifold  with solid lines is the fundamental manifold with the fit coefficients from \citet{zar08eqgal} corrected to V-band (see text).  The (blue) plane with dashed borders corresponds to $M_{1/2} = 3 L_{1/2}$, the mass-follows-light plane.   The data point color and point-type scheme matches that of the Figure \ref{fig:mlr3d} (or see \S \ref{sec:dat}). 
A rotating animation of this plot is available at \movieurl.}
\label{fig:mlr3dfman}
\end{figure*}

In Figure \ref{fig:mlr3dfman} we show this data set again in MRL space (as in Figure \ref{fig:mlr3d}), but we now overplot the fundamental manifold of \citet{zar06fman} in transparent green.  We use the fundamental manifold from \citet{zar08eqgal} Table 1, using the transforms from \S \ref{sec:mlrspace} to convert from fundamental plane space to MRL space.  We also adjust the luminosity from the value for I-band \citep[used in ][]{zar08eqgal} to V-band assuming all objects have V-I colors of typical E galaxies from \citet{fuk95}. While there will be an additional bias because $R_e$ for V and I bands will differ, this is likely small relative to the scatter and hence we disregard it.

Finally, we show the mean mass density of our data set as derived from the middle panel of Figure \ref{fig:mlr2d}. This derived simply by tilting the $\log(r_{1/2}) - \log(M_{1/2})$ relation to give density within $r_{1/2}$ instead of mass (i.e. residuals from the dashed-dotted line in the middle panel of Figure \ref{fig:mlr2d}, but with a different normalization fixed to standard density units).

\begin{figure*}[htb!]
\epsscale{1.15}

\plotone{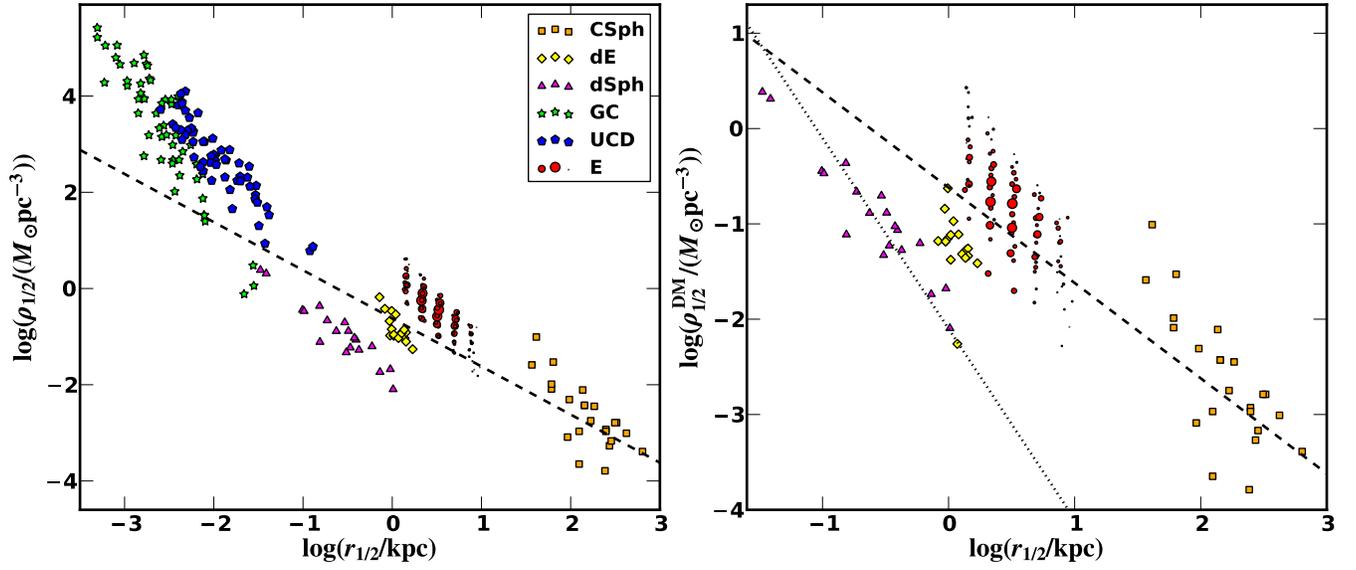}
\caption{Mean mass density within $r_{1/2}$ as derived from the $r_{1/2}-M_{1/2}$ relation (e.g. middle panel of Figure \ref{fig:mlr2d}). The left panel uses the raw dynamical mass $M_{1/2}$ to compute the density, while the right panel uses the mass $M_{1/2}^{\rm DM}$ for which the stellar contribution has been subtracted.  To guide the eye, we include the $\rho_{1/2} \propto r^{-1}$ relation (black dashed lines), and for the right panel, the isothermal case of $\rho_{1/2} \propto r^{-2}$ (black dotted line).  The color/symbol code maps to galaxy type as indicated, matching the scheme of Figure \ref{fig:fjplot}.}
\label{fig:densityplot}
\end{figure*}

\clearpage

\section{UCD Mass Estimates}
\label{apx:UCDs}

As discussed in \S \ref{sec:mlrspace}, UCDs present a puzzle in the MRL space. The most massive UCDs approach the fundamental curve, although with a gap that could potentially be a result of selection effects.  Regardless, the current sample of UCDs form a distinct group (with GCs) from the dSphs for the faintest/smallest objects.   Thus, if this sample of UCDs and dSphs are both galaxies, the scaling relations split into a dichotomy or bimodality at the faint end.  For this paper we have focused on the dSph side of this relation, but here we consider the UCDs in the profile matching context.  

\begin{figure}[htbp]
\plotone{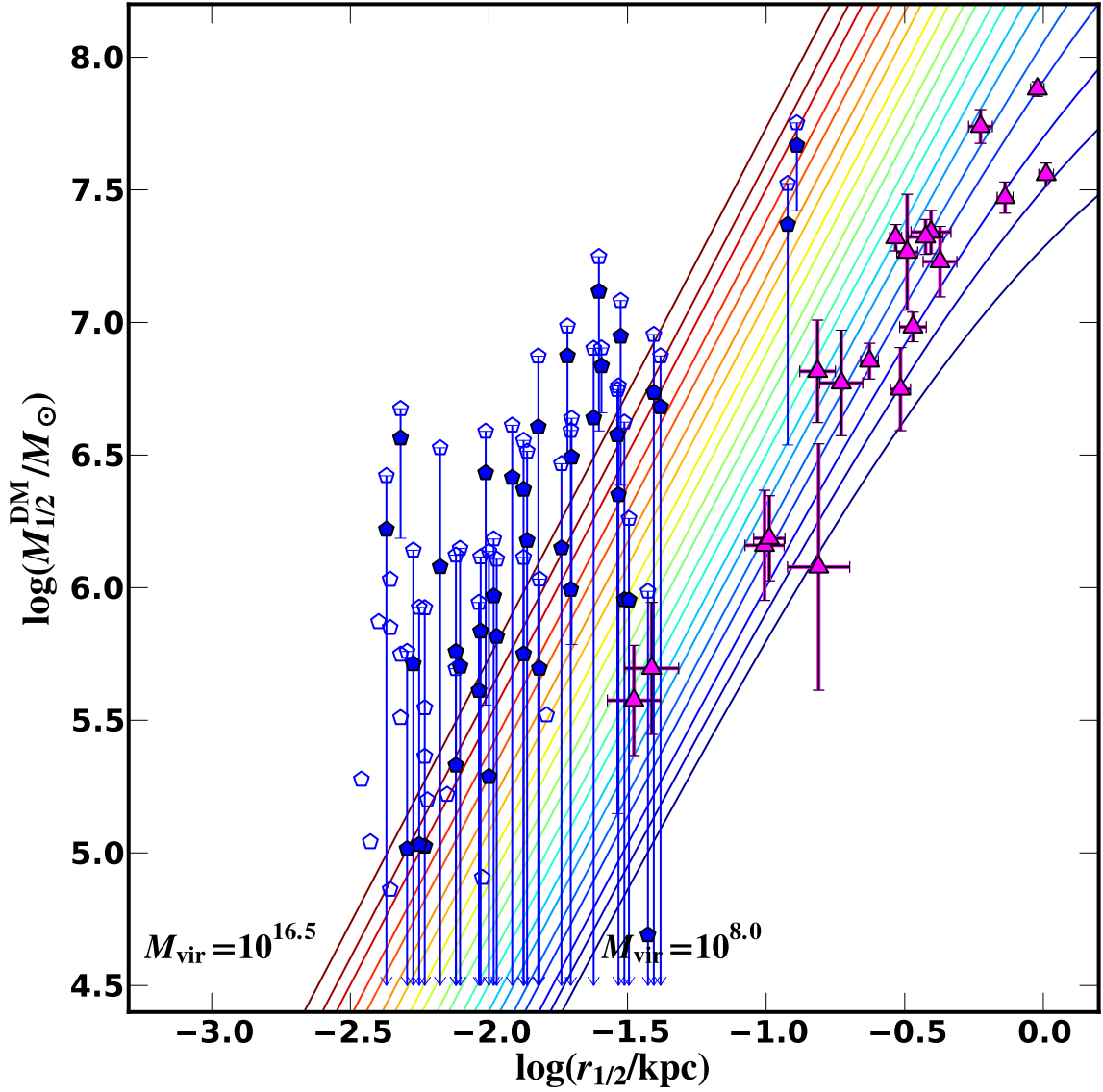}
\caption{Profile matching schematic for UCDs and dSphs, based on Figure \ref{fig:massplot}.  Open blue circles are UCDs with mass-to-light ratios ($\Upsilon$) of 1 assumed for computing $M_{1/2}^{\rm DM}$, while solid blue circles assume $\Upsilon=2$, and error mass error bars assume a factor of 2 uncertainty in $\Upsilon$. Upper limit error bars (extending to the bottom of the plot) correspond to those where the systematic uncertainty could result in a UCD fully consistent with its stellar mass implied from the luminosity.  Magenta triangles are dSph galaxies,with observational error bars.  The colored lines are the grid of Figure \ref{fig:massplot}, representing the mass profiles for a range of NFW halos.  Thus the densities of UCDs imply that they are either within the most-massive halos, or primarily purely stellar.}
\label{fig:UCDdSphmassplots}
\end{figure}

Figure \ref{fig:UCDdSphmassplots} is analogous to Figure \ref{fig:massplot}, but zoomed in on the faint end and with UCDs added.  Note that for UCDs, we determine the dark matter mass for the UCDs as  $M_{1/2}^{\rm DM} = M_{1/2}  -  L_{1/2} \Upsilon$, where an $\Upsilon$ is taken to be fixed at 1 (open circles) or 2 (filled circles).  For the latter, we also show error bars based on a possible factor of 2 systematic uncertainty in stellar models, based on the discussion in \S \ref{sec:err} for E galaxies. The dSph error bars are from \ref{fig:massplot}, based on the observational error bars in $r_{1/2}$ and $\sigma$.

The grid of NFW halos in Figure \ref{fig:UCDdSphmassplots} clearly shows that the implied dark matter densities for UCDs are most consistent with cluster-sized (or larger) dark matter halos ($M_{\rm vir} \gtrsim 10^{15}$) .  Taken at face value, this is impossible, as there are not enough of such halos where UCDs are found, and they would have clear kinematic effects on neighbors if UCDs had such massive halos.  A few possibilities might explain these large virial masses.  If UCDs do indeed have dark matter halos, baryonic contraction might boost their central densities (as described in \S \ref{sec:halomatch} for E galaxies).  However, given the extreme stellar densities and small sizes of UCDs (and hence short dynamical times), it seems unlikely that any baryonic contraction would be adiabatic.  Hence we cannot apply  adiabatic contraction as we have used to correct masses for the Es.  A baryonic contraction model appropriate for UCDs could be used in the same way, although we do not do such a correction as such a model does not yet exist.

An alternative possibility is that the stellar population estimates are systematically in error.  The error bars in \ref{fig:UCDdSphmassplots} imply that such errors could explain most (possibly all) UCDs as entirely stellar objects--this corresponds to those where the error bars are upper limits.  Alternatively, they may have dark matter halos with much smaller virial masses, but without better stellar population models, there is no way to tell the difference.  

Thus, while UCDs are possibly consistent with lying inside dark matter halos, this implies a dichotomy in galaxy formation as well as being impossible to explain with standard $\Lambda$CDM dark matter halos.  We therefore favor the simplest view that they are purely stellar systems.

\end{document}